\magnification=\magstep1
\def\firstpage{1}
\pageno=\firstpage
\font\fiverm=cmr5
\font\sevenrm=cmr7

\font\eightrm=cmr8
\font\eightbf=cmbx8
\font\ninerm=cmr9
\font\ninebf=cmbx9
\font\tenbf=cmbx10
\font\twelvebf=cmbx12
%

%

%

%
%
\newskip\ttglue
\font\fiverm=cmr5
\font\fivei=cmmi5
\font\fivesy=cmsy5
\font\fivebf=cmbx5
\font\sixrm=cmr6
\font\sixi=cmmi6
\font\sixsy=cmsy6
\font\sixbf=cmbx6
\font\sevenrm=cmr7
\font\eightrm=cmr8
\font\eighti=cmmi8
\font\eightsy=cmsy8
\font\eightit=cmti8
\font\eightsl=cmsl8
\font\eighttt=cmtt8
\font\eightbf=cmbx8
\font\ninerm=cmr9
\font\ninei=cmmi9
\font\ninesy=cmsy9
\font\nineit=cmti9
\font\ninesl=cmsl9
\font\ninett=cmtt9
\font\ninebf=cmbx9
\font\twelverm=cmr12
\font\twelvei=cmmi12
\font\twelvesy=cmsy12
\font\twelveit=cmti12
\font\twelvesl=cmsl12
\font\twelvett=cmtt12
\font\twelvebf=cmbx12


\def\eightpoint{\def\rm{\fam0\eightrm}
  \textfont0=\eightrm \scriptfont0=\sixrm \scriptscriptfont0=\fiverm
  \textfont1=\eighti  \scriptfont1=\sixi  \scriptscriptfont1=\fivei
  \textfont2=\eightsy  \scriptfont2=\sixsy  \scriptscriptfont2=\fivesy
  \textfont3=\tenex  \scriptfont3=\tenex  \scriptscriptfont3=\tenex
  \textfont\itfam=\eightit  \def\it{\fam\itfam\eightit}
  \textfont\slfam=\eightsl  \def\sl{\fam\slfam\eightsl}
  \textfont\ttfam=\eighttt  \def\tt{\fam\ttfam\eighttt}
  \textfont\bffam=\eightbf  \scriptfont\bffam=\sixbf
    \scriptscriptfont\bffam=\fivebf  \def\bf{\fam\bffam\eightbf}
  \tt  \ttglue=.5em plus.25em minus.15em
  \normalbaselineskip=9pt
  \setbox\strutbox=\hbox{\vrule height7pt depth2pt width0pt}
  \let\sc=\sixrm  \let\big=\eightbig \normalbaselines\rm}

\def\eightbig#1{{\hbox{$\textfont0=\ninerm\textfont2=\ninesy
        \left#1\vbox to6.5pt{}\right.$}}}


\def\ninepoint{\def\rm{\fam0\ninerm}
  \textfont0=\ninerm \scriptfont0=\sixrm \scriptscriptfont0=\fiverm
  \textfont1=\ninei  \scriptfont1=\sixi  \scriptscriptfont1=\fivei
  \textfont2=\ninesy  \scriptfont2=\sixsy  \scriptscriptfont2=\fivesy
  \textfont3=\tenex  \scriptfont3=\tenex  \scriptscriptfont3=\tenex
  \textfont\itfam=\nineit  \def\it{\fam\itfam\nineit}
  \textfont\slfam=\ninesl  \def\sl{\fam\slfam\ninesl}
  \textfont\ttfam=\ninett  \def\tt{\fam\ttfam\ninett}
  \textfont\bffam=\ninebf  \scriptfont\bffam=\sixbf
    \scriptscriptfont\bffam=\fivebf  \def\bf{\fam\bffam\ninebf}
  \tt  \ttglue=.5em plus.25em minus.15em
  \normalbaselineskip=11pt
  \setbox\strutbox=\hbox{\vrule height8pt depth3pt width0pt}
  \let\sc=\sevenrm  \let\big=\ninebig \normalbaselines\rm}

\def\ninebig#1{{\hbox{$\textfont0=\tenrm\textfont2=\tensy
        \left#1\vbox to7.25pt{}\right.$}}}


\def\twelvepoint{\def\rm{\fam0\twelverm}
  \textfont0=\twelverm \scriptfont0=\eightrm \scriptscriptfont0=\sixrm
  \textfont1=\twelvei  \scriptfont1=\eighti  \scriptscriptfont1=\sixi
  \textfont2=\twelvesy  \scriptfont2=\eightsy  \scriptscriptfont2=\sixsy
  \textfont3=\tenex  \scriptfont3=\tenex  \scriptscriptfont3=\tenex
  \textfont\itfam=\twelveit  \def\it{\fam\itfam\twelveit}
  \textfont\slfam=\twelvesl  \def\sl{\fam\slfam\twelvesl}
  \textfont\ttfam=\twelvett  \def\tt{\fam\ttfam\twelvett}
  \textfont\bffam=\twelvebf  \scriptfont\bffam=\eightbf
    \scriptscriptfont\bffam=\sixbf  \def\bf{\fam\bffam\twelvebf}
  \tt  \ttglue=.5em plus.25em minus.15em
  \normalbaselineskip=11pt
  \setbox\strutbox=\hbox{\vrule height8pt depth3pt width0pt}
  \let\sc=\sevenrm  \let\big=\twelvebig \normalbaselines\rm}

\def\twelvebig#1{{\hbox{$\textfont0=\tenrm\textfont2=\tensy
        \left#1\vbox to7.25pt{}\right.$}}}
\catcode`\@=11
%

\def\undefine#1{\let#1\undefined}
\def\newsymbol#1#2#3#4#5{\let\next@\relax
 \ifnum#2=\@ne\let\next@\msafam@\else
 \ifnum#2=\tw@\let\next@\msbfam@\fi\fi
 \mathchardef#1="#3\next@#4#5}
\def\mathhexbox@#1#2#3{\relax
 \ifmmode\mathpalette{}{\m@th\mathchar"#1#2#3}%
 \else\leavevmode\hbox{$\m@th\mathchar"#1#2#3$}\fi}
\def\hexnumber@#1{\ifcase#1 0\or 1\or 2\or 3\or 4\or 5\or 6\or 7\or 8\or
 9\or A\or B\or C\or D\or E\or F\fi}

\font\tenmsa=msam10
\font\sevenmsa=msam7
\font\fivemsa=msam5
\newfam\msafam
\textfont\msafam=\tenmsa
\scriptfont\msafam=\sevenmsa
\scriptscriptfont\msafam=\fivemsa
\edef\msafam@{\hexnumber@\msafam}
\mathchardef\dabar@"0\msafam@39
\def\dashrightarrow{\mathrel{\dabar@\dabar@\mathchar"0\msafam@4B}}
\def\dashleftarrow{\mathrel{\mathchar"0\msafam@4C\dabar@\dabar@}}

\def\ulcorner{\delimiter"4\msafam@70\msafam@70 }
\def\urcorner{\delimiter"5\msafam@71\msafam@71 }
\def\llcorner{\delimiter"4\msafam@78\msafam@78 }
\def\lrcorner{\delimiter"5\msafam@79\msafam@79 }
\def\yen{{\mathhexbox@\msafam@55}}
\def\checkmark{{\mathhexbox@\msafam@58}}
\def\circledR{{\mathhexbox@\msafam@72}}
\def\maltese{{\mathhexbox@\msafam@7A}}

\font\tenmsb=msbm10
\font\sevenmsb=msbm7
\font\fivemsb=msbm5
\newfam\msbfam
\textfont\msbfam=\tenmsb
\scriptfont\msbfam=\sevenmsb
\scriptscriptfont\msbfam=\fivemsb
\edef\msbfam@{\hexnumber@\msbfam}
\def\Bbb#1{{\fam\msbfam\relax#1}}
\def\widehat#1{\setbox\z@\hbox{$\m@th#1$}%
 \ifdim\wd\z@>\tw@ em\mathaccent"0\msbfam@5B{#1}%
 \else\mathaccent"0362{#1}\fi}
\def\widetilde#1{\setbox\z@\hbox{$\m@th#1$}%
 \ifdim\wd\z@>\tw@ em\mathaccent"0\msbfam@5D{#1}%
 \else\mathaccent"0365{#1}\fi}
\font\teneufm=eufm10
\font\seveneufm=eufm7
\font\fiveeufm=eufm5
\newfam\eufmfam
\textfont\eufmfam=\teneufm
\scriptfont\eufmfam=\seveneufm
\scriptscriptfont\eufmfam=\fiveeufm

\catcode`\@=11
\newsymbol\boxdot 1200
\newsymbol\boxplus 1201
\newsymbol\boxtimes 1202
\newsymbol\square 1003
\newsymbol\blacksquare 1004
\newsymbol\centerdot 1205
\newsymbol\lozenge 1006
\newsymbol\blacklozenge 1007
\newsymbol\circlearrowright 1308
\newsymbol\circlearrowleft 1309
\undefine\rightleftharpoons
\newsymbol\rightleftharpoons 130A
\newsymbol\leftrightharpoons 130B
\newsymbol\boxminus 120C
\newsymbol\Vdash 130D
\newsymbol\Vvdash 130E
\newsymbol\vDash 130F
\newsymbol\twoheadrightarrow 1310
\newsymbol\twoheadleftarrow 1311
\newsymbol\leftleftarrows 1312
\newsymbol\rightrightarrows 1313
\newsymbol\upuparrows 1314
\newsymbol\downdownarrows 1315
\newsymbol\upharpoonright 1316
 
\newsymbol\downharpoonright 1317
\newsymbol\upharpoonleft 1318
\newsymbol\downharpoonleft 1319
\newsymbol\rightarrowtail 131A
\newsymbol\leftarrowtail 131B
\newsymbol\leftrightarrows 131C
\newsymbol\rightleftarrows 131D
\newsymbol\Lsh 131E
\newsymbol\Rsh 131F
\newsymbol\rightsquigarrow 1320
\newsymbol\leftrightsquigarrow 1321
\newsymbol\looparrowleft 1322
\newsymbol\looparrowright 1323
\newsymbol\circeq 1324
\newsymbol\succsim 1325
\newsymbol\gtrsim 1326
\newsymbol\gtrapprox 1327
\newsymbol\multimap 1328
\newsymbol\therefore 1329
\newsymbol\because 132A
\newsymbol\doteqdot 132B
 
\newsymbol\triangleq 132C
\newsymbol\precsim 132D
\newsymbol\lesssim 132E
\newsymbol\lessapprox 132F
\newsymbol\eqslantless 1330
\newsymbol\eqslantgtr 1331
\newsymbol\curlyeqprec 1332
\newsymbol\curlyeqsucc 1333
\newsymbol\preccurlyeq 1334
\newsymbol\leqq 1335
\newsymbol\leqslant 1336
\newsymbol\lessgtr 1337
\newsymbol\backprime 1038
\newsymbol\risingdotseq 133A
\newsymbol\fallingdotseq 133B
\newsymbol\succcurlyeq 133C
\newsymbol\geqq 133D
\newsymbol\geqslant 133E
\newsymbol\gtrless 133F
\newsymbol\sqsubset 1340
\newsymbol\sqsupset 1341
\newsymbol\vartriangleright 1342
\newsymbol\vartriangleleft 1343
\newsymbol\trianglerighteq 1344
\newsymbol\trianglelefteq 1345
\newsymbol\bigstar 1046
\newsymbol\between 1347
\newsymbol\blacktriangledown 1048
\newsymbol\blacktriangleright 1349
\newsymbol\blacktriangleleft 134A
\newsymbol\vartriangle 134D
\newsymbol\blacktriangle 104E
\newsymbol\triangledown 104F
\newsymbol\eqcirc 1350
\newsymbol\lesseqgtr 1351
\newsymbol\gtreqless 1352
\newsymbol\lesseqqgtr 1353
\newsymbol\gtreqqless 1354
\newsymbol\Rrightarrow 1356
\newsymbol\Lleftarrow 1357
\newsymbol\veebar 1259
\newsymbol\barwedge 125A
\newsymbol\doublebarwedge 125B
\undefine\angle
\newsymbol\angle 105C
\newsymbol\measuredangle 105D
\newsymbol\sphericalangle 105E
\newsymbol\varpropto 135F
\newsymbol\smallsmile 1360
\newsymbol\smallfrown 1361
\newsymbol\Subset 1362
\newsymbol\Supset 1363
\newsymbol\Cup 1264
 
\newsymbol\Cap 1265
 
\newsymbol\curlywedge 1266
\newsymbol\curlyvee 1267
\newsymbol\leftthreetimes 1268
\newsymbol\rightthreetimes 1269
\newsymbol\subseteqq 136A
\newsymbol\supseteqq 136B
\newsymbol\bumpeq 136C
\newsymbol\Bumpeq 136D
\newsymbol\lll 136E
 
\newsymbol\ggg 136F
 
\newsymbol\circledS 1073
\newsymbol\pitchfork 1374
\newsymbol\dotplus 1275
\newsymbol\backsim 1376
\newsymbol\backsimeq 1377
\newsymbol\complement 107B
\newsymbol\intercal 127C
\newsymbol\circledcirc 127D
\newsymbol\circledast 127E
\newsymbol\circleddash 127F
\newsymbol\lvertneqq 2300
\newsymbol\gvertneqq 2301
\newsymbol\nleq 2302
\newsymbol\ngeq 2303
\newsymbol\nless 2304
\newsymbol\ngtr 2305
\newsymbol\nprec 2306
\newsymbol\nsucc 2307
\newsymbol\lneqq 2308
\newsymbol\gneqq 2309
\newsymbol\nleqslant 230A
\newsymbol\ngeqslant 230B
\newsymbol\lneq 230C
\newsymbol\gneq 230D
\newsymbol\npreceq 230E
\newsymbol\nsucceq 230F
\newsymbol\precnsim 2310
\newsymbol\succnsim 2311
\newsymbol\lnsim 2312
\newsymbol\gnsim 2313
\newsymbol\nleqq 2314
\newsymbol\ngeqq 2315
\newsymbol\precneqq 2316
\newsymbol\succneqq 2317
\newsymbol\precnapprox 2318
\newsymbol\succnapprox 2319
\newsymbol\lnapprox 231A
\newsymbol\gnapprox 231B
\newsymbol\nsim 231C
\newsymbol\ncong 231D
\newsymbol\diagup 201E
\newsymbol\diagdown 201F
\newsymbol\varsubsetneq 2320
\newsymbol\varsupsetneq 2321
\newsymbol\nsubseteqq 2322
\newsymbol\nsupseteqq 2323
\newsymbol\subsetneqq 2324
\newsymbol\supsetneqq 2325
\newsymbol\varsubsetneqq 2326
\newsymbol\varsupsetneqq 2327
\newsymbol\subsetneq 2328
\newsymbol\supsetneq 2329
\newsymbol\nsubseteq 232A
\newsymbol\nsupseteq 232B
\newsymbol\nparallel 232C
\newsymbol\nmid 232D
\newsymbol\nshortmid 232E
\newsymbol\nshortparallel 232F
\newsymbol\nvdash 2330
\newsymbol\nVdash 2331
\newsymbol\nvDash 2332
\newsymbol\nVDash 2333
\newsymbol\ntrianglerighteq 2334
\newsymbol\ntrianglelefteq 2335
\newsymbol\ntriangleleft 2336
\newsymbol\ntriangleright 2337
\newsymbol\nleftarrow 2338
\newsymbol\nrightarrow 2339
\newsymbol\nLeftarrow 233A
\newsymbol\nRightarrow 233B
\newsymbol\nLeftrightarrow 233C
\newsymbol\nleftrightarrow 233D
\newsymbol\divideontimes 223E
\newsymbol\varnothing 203F
\newsymbol\nexists 2040
\newsymbol\Finv 2060
\newsymbol\Game 2061
\newsymbol\mho 2066
\newsymbol\eth 2067
\newsymbol\eqsim 2368
\newsymbol\beth 2069
\newsymbol\gimel 206A
\newsymbol\daleth 206B
\newsymbol\lessdot 236C
\newsymbol\gtrdot 236D
\newsymbol\ltimes 226E
\newsymbol\rtimes 226F
\newsymbol\shortmid 2370
\newsymbol\shortparallel 2371
\newsymbol\smallsetminus 2272
\newsymbol\thicksim 2373
\newsymbol\thickapprox 2374
\newsymbol\approxeq 2375
\newsymbol\succapprox 2376
\newsymbol\precapprox 2377
\newsymbol\curvearrowleft 2378
\newsymbol\curvearrowright 2379
\newsymbol\digamma 207A
\newsymbol\varkappa 207B
\newsymbol\Bbbk 207C
\newsymbol\hslash 207D
\undefine\hbar
\newsymbol\hbar 207E
\newsymbol\backepsilon 237F

%
\newcount\marknumber	\marknumber=1
\newcount\countdp \newcount\countwd \newcount\countht
%
%
\ifx\pdfoutput\undefined
\def\rgboo#1{}
\def\postscript#1{\special{" #1}}		
\postscript{
	/bd {bind def} bind def
	/fsd {findfont exch scalefont def} bd
	/sms {setfont moveto show} bd
	/ms {moveto show} bd
	/pdfmark where		
	{pop} {userdict /pdfmark /cleartomark load put} ifelse
	[ /PageMode /UseOutlines		
	/DOCVIEW pdfmark}
\def\bookmark#1#2{\postscript{		
	[ /Dest /MyDest\the\marknumber /View [ /XYZ null null null ] /DEST pdfmark
	[ /Title (#2) /Count #1 /Dest /MyDest\the\marknumber /OUT pdfmark}%
	\advance\marknumber by1}
\def\pdfclink#1#2#3{%
	\hskip-.25em\setbox0=\hbox{#2}%
		\countdp=\dp0 \countwd=\wd0 \countht=\ht0%
		\divide\countdp by65536 \divide\countwd by65536%
			\divide\countht by65536%
		\advance\countdp by1 \advance\countwd by1%
			\advance\countht by1%
		\def\linkdp{\the\countdp} \def\linkwd{\the\countwd}%
			\def\linkht{\the\countht}%
	\postscript{
		[ /Rect [ -1.5 -\linkdp.0 0\linkwd.0 0\linkht.5 ]
		/Border [ 0 0 0 ]
		/Action << /Subtype /URI /URI (#3) >>
		/Subtype /Link
		/ANN pdfmark}{\rgb{#1}{#2}}}
%
%
\else
\def\rgboo#1{\pdfliteral{#1 rg #1 RG}}
\pdfcatalog{/PageMode /UseOutlines}		
\def\bookmark#1#2{
	\pdfdest num \marknumber xyz
	\pdfoutline goto num \marknumber count #1 {#2}
	\advance\marknumber by1}
\def\pdfklink#1#2{%
	\noindent\pdfstartlink user
		{/Subtype /Link
		/Border [ 0 0 0 ]
		/A << /S /URI /URI (#2) >>}{\rgb{1 0 0}{#1}}%
	\pdfendlink}
\fi

\def\rgbo#1#2{\rgboo{#1}#2\rgboo{0 0 0}}
\def\rgb#1#2{\mark{#1}\rgbo{#1}{#2}\mark{0 0 0}}
\def\pdfklink#1#2{\pdfclink{1 0 0}{#1}{#2}}
\def\pdflink#1{\pdfklink{#1}{#1}}
%
%
\newcount\seccount  
\newcount\subcount  
\newcount\clmcount  
\newcount\equcount  
\newcount\refcount  
\newcount\demcount  
\newcount\execount  
\newcount\procount  
\seccount=0
\equcount=1
\clmcount=1
\subcount=1
\refcount=1
\demcount=0
\execount=0
\procount=0
%

\def\proofof(#1){\medskip\noindent{\bf Proof of \csname c#1\endcsname.\ }}

\def\references{\bigskip\noindent\hbox{\bf References}\medskip
                \ifx\pdflink\undefined\else\bookmark{0}{References}\fi}
\def\addref#1{\expandafter\xdef\csname r#1\endcsname{\number\refcount}
    \global\advance\refcount by 1}

\def\nextremark #1\par{\item{$\circ$} #1}
\def\firstremark #1\par{\bigskip\noindent{\bf Remarks.}
     \smallskip\nextremark #1\par}
\def\abstract#1\par{{\baselineskip=10pt
    \eightpoint\narrower\noindent{\eightbf Abstract.} #1\par}}
%
\def\equtag#1{\expandafter\xdef\csname e#1\endcsname{(\number\seccount.\number\equcount)}
              \global\advance\equcount by 1}
\def\equation(#1){\equtag{#1}\eqno\csname e#1\endcsname}
\def\equ(#1){\hskip-0.03em\csname e#1\endcsname}
%
\def\clmtag#1#2{\expandafter\xdef\csname cn#2\endcsname{\number\seccount.\number\clmcount}
                \expandafter\xdef\csname c#2\endcsname{#1~\number\seccount.\number\clmcount}
                \global\advance\clmcount by 1}
\def\claim #1(#2) #3\par{\clmtag{#1}{#2}
    \vskip.1in\medbreak\noindent
    {\bf \csname c#2\endcsname .\ }{\sl #3}\par
    \ifdim\lastskip<\medskipamount
    \removelastskip\penalty55\medskip\fi}
\def\clm(#1){\csname c#1\endcsname}
\def\clmno(#1){\csname cn#1\endcsname}
%
\def\sectag#1{\global\advance\seccount by 1
              \expandafter\xdef\csname sectionname\endcsname{\number\seccount. #1}
              \equcount=1 \clmcount=1 \subcount=1 \execount=0 \procount=0}
\def\section#1\par{\vskip0pt plus.1\vsize\penalty-40
    \vskip0pt plus -.1\vsize\bigskip\bigskip
    \sectag{#1}
    \message{\sectionname}\leftline{\twelvebf\sectionname} 
    \nobreak\smallskip\noindent
    \ifx\pdflink\undefined
    \else
      \bookmark{0}{\sectionname}
    \fi}
%
\def\subtag#1{\expandafter\xdef\csname subsectionname\endcsname{\number\seccount.\number\subcount. #1}
              \global\advance\subcount by 1}
\def\subsection#1\par{\vskip0pt plus.05\vsize\penalty-20
    \vskip0pt plus -.05\vsize\medskip\medskip
    \subtag{#1}
    \message{\subsectionname}\leftline{\tenbf\subsectionname}
    \nobreak\smallskip\noindent
    \ifx\pdflink\undefined
    \else
      \bookmark{0}{.... \subsectionname}  
    \fi}
%
\def\demtag#1#2{\global\advance\demcount by 1
              \expandafter\xdef\csname de#2\endcsname{#1~\number\demcount}}
\def\demo #1(#2) #3\par{
  \demtag{#1}{#2}
  \vskip.1in\medbreak\noindent
  {\bf #1 \number\demcount.\enspace}
  {\rm #3}\par
  \ifdim\lastskip<\medskipamount
  \removelastskip\penalty55\medskip\fi}
\def\dem(#1){\csname de#1\endcsname}
%
\def\exetag#1{\global\advance\execount by 1
              \expandafter\xdef\csname ex#1\endcsname{Exercise~\number\seccount.\number\execount}}
\def\exercise(#1) #2\par{
  \exetag{#1}
  \vskip.1in\medbreak\noindent
  {\bf Exercise \number\execount.}
  {\rm #2}\par
  \ifdim\lastskip<\medskipamount
  \removelastskip\penalty55\medskip\fi}
\def\exe(#1){\csname ex#1\endcsname}
%
\def\protag#1{\global\advance\procount by 1
              \expandafter\xdef\csname pr#1\endcsname{\number\seccount.\number\procount}}
\def\problem(#1) #2\par{
  \ifnum\procount=0
    \parskip=6pt
    \vbox{\bigskip\centerline{\bf Problems \number\seccount}\nobreak\medskip}
  \fi
  \protag{#1}
  \item{\number\procount.} #2}
\def\pro(#1){Problem \csname pr#1\endcsname}
%
%
%
\def\rightheadline{\hfil}
\def\leftheadline{\sevenrm\hfil HANS KOCH\hfil}
\headline={\ifnum\pageno=\firstpage\hfil\else
\ifodd\pageno{{\fiverm\rightheadline}\number\pageno}
\else{\number\pageno\fiverm\leftheadline}\fi\fi}
\footline={\ifnum\pageno=\firstpage\hss\tenrm\folio\hss\else\hss\fi}

\let\cl=\centerline

\let\eps=\varepsilon
\let\sss=\scriptscriptstyle

\def\AA{{\cal A}}
\def\BB{{\cal B}}
\def\CC{{\cal C}}

\def\EE{{\cal E}}

\def\NN{{\cal N}}
\def\OO{{\cal O}}
\def\PP{{\cal P}}

\def\SS{{\cal S}}

\def\XX{{\cal X}}
\def\YY{{\cal Y}}

\def\ssN{{\sss N}}

\def\id{{\rm I}}

\def\tr{\mathop{\rm tr}\nolimits}
\def\det{\mathop{\rm det}\nolimits}

\def\diag{\mathop{\rm diag}\nolimits}

%
\newfam\dsfam
\def\mathds #1{{\fam\dsfam\tends #1}}

\font\tends=dsrom10
\font\eightds=dsrom8
\textfont\dsfam=\tends
\scriptfont\dsfam=\eightds
%

\def\integer{{\mathds Z}}

\def\real{{\mathds R}}
\def\complex{{\mathds C}}

\def\torus{{\Bbb T}}

\def\oo{{\scriptstyle\OO}}
\def\bdot{\hbox{\bf .}}
\def\bcomma{\hbox{\bf ,}}
\def\defeq{\mathrel{\mathop=^{\sss\rm def}}}
\def\half{{1\over 2}}

\def\thalf{{\textstyle\half}}

\def\twovec#1#2{\left[\matrix{#1\cr#2\cr}\right]}

\def\twomat#1#2#3#4{\left[\matrix{#1&#2\cr#3&#4\cr}\right]}

%

%

%

%

\input miniltx

\ifx\pdfoutput\undefined
  \def\Gin@driver{dvips.def}  
\else
  \def\Gin@driver{pdftex.def} 
\fi

\input graphicx.sty
\resetatcatcode
%
\newdimen\savedparindent
\savedparindent=\parindent
\font\tenamsb=msbm10 \font\sevenamsb=msbm7 \font\fiveamsb=msbm5
\newfam\bbfam
\textfont\bbfam=\tenamsb
\scriptfont\bbfam=\sevenamsb
\scriptscriptfont\bbfam=\fiveamsb

\def\AM{{\ninerm AM~}}
\def\RG{{\ninerm RG~}}

\def\buF{{\hbox{\teneufm F}}}

\def\buM{{\hbox{\teneufm M}}}
\def\buN{{\hbox{\teneufm N}}}
\def\buR{{\hbox{\teneufm R}}}

\font\fourrm=cmr5 at 4pt
\def\srmF{{\hbox{\fiverm F}}}
\def\srmG{{\hbox{\fiverm G}}}
\def\ssrmF{{\hbox{\fourrm F}}}
\def\ssrmG{{\hbox{\fourrm G}}}
\font\tenib=cmmib10
\def\bma{{\hbox{\tenib a}}}

\def\bms{{\hbox{\tenib s}}}

\def\col{{:\hskip3pt}}
\def\scol{{;\hskip3pt}}
\def\rot{\mathop{\rm rot}\nolimits}
\def\mod{\mathop{\rm mod}\nolimits}
\def\gcd{\mathop{\rm gcd}\nolimits}

\def\idmat{{\bf 1}}

\def\rmSL{{\rm SL}}
\def\rmSU{{\rm SU}}
\def\rmPSL{{\rm PSL}}

\def\ssN{{\sss N}}

\def\old{{\sss{\rm old}}}
\def\new{{\sss{\rm new}}}

\def\tinyskip{\hskip.7pt}
\def\nominus{\phantom{-}}

\def\oo{{\scriptstyle\OO}}

\def\bdot{\hbox{\bf .}}
\def\hdots{\line{\leaders\hbox to 0.5em{\hss .\hss}\hfil}}

\def\sfrac#1#2{\hbox{\raise2.2pt\hbox{$\scriptstyle#1$}\hskip-1.2pt
   {$\scriptstyle/$}\hskip-0.9pt\lower2.2pt\hbox{$\scriptstyle#2$}\hskip1.0pt}}
\def\shalf{\sfrac{1}{2}}

\def\stwomat#1#2#3#4{{\eightpoint\left[\matrix{#1&#2\cr#3&#4\cr}\right]}}

\def\today{\ifcase\month\or
January\or February\or March\or April\or May\or June\or
July\or August\or September\or October\or November\or December\fi
\space\number\day, \number\year}
\addref{Harp}
\addref{Hof}
\addref{Kada}
\addref{McK}
\addref{BeSi}
\addref{JoMo}
\addref{AvSi}
\addref{OsKi}
\addref{DecJ}
\addref{Thou}
\addref{MR}
\addref{LaWi}
\addref{Lasti}
\addref{BES}
\addref{Lastii}
\addref{KeSa}
\addref{Stirn}
\addref{GJLS}
\addref{RuPi}
\addref{Jito}
\addref{MeOsWi}
\addref{OsAv}
\addref{RK}
\addref{MeOs}
\addref{AvKr}
\addref{Dama}
\addref{GoSch}
\addref{AvDa}
\addref{AK}
\addref{AKfhn}
\addref{Satij}
\addref{Koch}
\addref{OEIS}
\addref{KoKo}
\addref{Files}
\addref{Ada}
\addref{Gnat}
\addref{IEEE}
\addref{MPFR}
\def\leftheadline{\fiverm\hfil H.~KOCH\hfil\today}
\def\rightheadline{\sevenrm\hfil Renormalization for the AM operator\hfil}
%
\cl{{\twelvebf Golden mean renormalization}}
\cl{{\twelvebf for the almost Mathieu operator and related skew products}}
\bigskip

\cl{
Hans Koch
\footnote{$^1$}
{\eightpoint\hskip-2.7em
Department of Mathematics, The University of Texas at Austin,
Austin, TX 78712.}
}

\bigskip
\abstract
Considering $\rmSL(2,{\scriptstyle\real})$ skew-product maps
over circle rotations, we prove that
a renormalization transformation associated with the golden mean $\alpha_\ast$
has a nontrivial periodic orbit of length $3$.
We also present some numerical results,
including evidence that this period $3$ describes
scaling properties of the Hofstadter butterfly
near the top of the spectrum at $\alpha_\ast$,
and scaling properties of the generalized eigenfunction for this energy.

\section Introduction

We consider a renormalization transformation
that arises in the study of the spectrum of
Schr\"odinger operators
$$
(H^\alpha u)_n=u_{n+1}+u_{n-1}+V(x_n)u_n\,,\qquad n\in\integer\,,
\equation(SchroedingerOp)
$$
acting on sequences $u\in\ell^2(\integer)$.
Here, $V$ is a suitable potential,
and $x_n=x_0+n\alpha$ for some given real number $\alpha$.
Potentials for which $n\mapsto V(x_n)$ is quasiperiodic
lead to interesting spectra
and have attracted considerable attention.
The equation $H^\alpha u=Eu$
for an eigenvector or generalized eigenvector of $H^\alpha$
can be written as
$$
\twovec{u_{n+1}}{u_n}=A(x_n)\twovec{u_n}{u_{n-1}}\,,\qquad
A(x)=\twomat{E-V(x)}{-1}{1}{0}\,.
\equation(SchroedingerCocycle)
$$

The motivating example for the work presented here is
the almost Mathieu ({\ninerm AM}) operator, which corresponds to
a potential $V(x)=2\lambda\cos(2\pi(x+\xi))$.
Two reviews can be found in [\rLastii,\rDama].
By adding $\shalf$ to $\xi$, if necessary,
we may assume that $\lambda\ge 0$.
A quantity of interest here is the rotation number
$$
\rot(\alpha,E)=\lim_{N\to\infty}{\Sigma_\ssN(\alpha,E)\over 2N}\,,
\equation(rotDef)
$$
where $\Sigma_\ssN(\alpha,E)$ denotes the
number of sign changes of a nontrivial solution $n\mapsto u_n$,
as $n$ ranges from $1$ to $N$.
For any fixed value of $x_0+\xi$, the rotation number $\rot(\alpha,E)$
is independent of $u$ and depends continuously on $\alpha$ and $E$.
If $\alpha$ is irrational, then $\rot(\alpha,E)$ is independent
of the choice of $x_0+\xi$ as well, by ergodicity.
For proof of these and other properties (mentioned below)
of the rotation number, we refer to [\rJoMo,\rAvSi,\rDecJ].

The \AM Hamiltonian $H^\alpha$ is a ``reduced'' form
of the Hofstadter Hamiltonian [\rHarp,\rHof],
which describes Bloch electrons moving on $\integer^2$,
under the influence a magnetic flux $2\pi\alpha$ through each unit cell.
For $\lambda<1$ the system is conducting (purely ac spectrum),
and for $\lambda>1$ it is insulating (purely pp spectrum),
for almost every value of $\alpha$ and $x_0+\xi$.
For details, including proofs and references, see [\rJito].
The Hofstadter Hamiltonian has an obvious duality transformation,
which corresponds to replacing $\lambda$ by $\lambda^{-1}$
and $E$ by $\lambda^{-1}E$.
In the self-dual case $\lambda=1$,
the spectrum of $H^\alpha$ is included in the interval $[-4,4]$,
and when plotted as a function of $\alpha\in[0,1]$,
it is known as the Hofstadter butterfly [\rHof].
It has zero Lebesgue measure [\rLasti]
and interesting topological properties [\rOsAv].
The spectrum itself is purely singular-continuous [\rGJLS],
for almost every value of $\alpha$ and $x_0+\xi$.

The Hofstadter butterfly is symmetric with respect
to the reflections $\alpha\mapsto 1-\alpha$ and $E\mapsto-E$.
The positive-energy part is shown in Figure 1.
The solid regions represent gaps in the spectrum,
which are open intervals for fixed $\alpha$;
and their colors encode the so-called gap index $k\in\integer$.
To be more precise, the function $\alpha\mapsto\rot(\alpha,E)$
is constant on the gap with index $k$, where it satisfies
$$
2\rot(\alpha,E)\equiv k\alpha\qquad(\mod 1)\,.
\equation(gapIndex)
$$
The left hand side of this congruence
can also be identified with the integrated density of states
[\rBeSi,\rAvSi,\rBES,\rGoSch],
which makes \equ(gapIndex) a purely spectral relation.

\vskip 0.6cm
\hbox{\hskip0.5cm
\includegraphics[height=6cm,width=12cm]{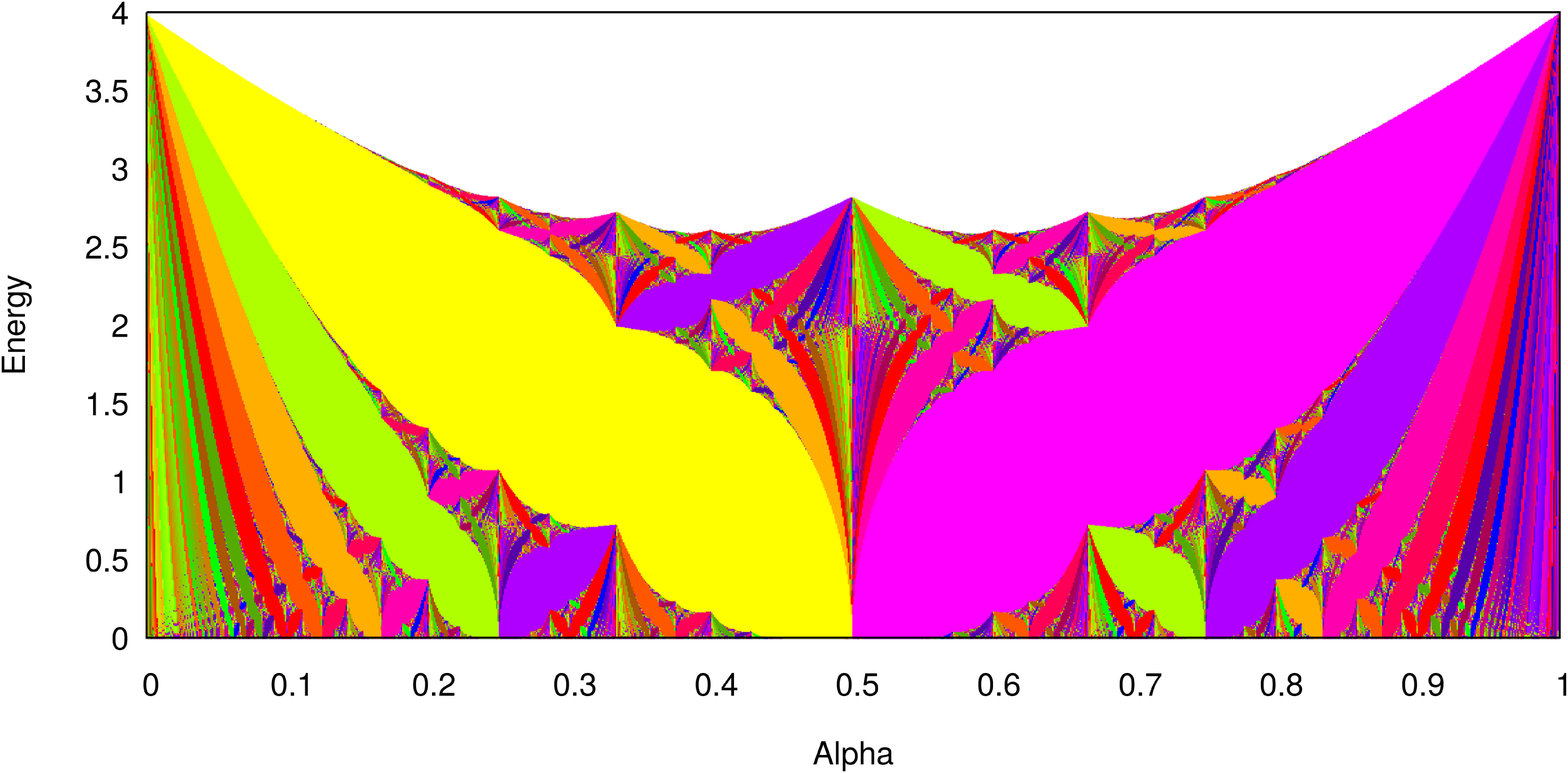}
}
\vskip 0.0cm
\leftline{\hskip 1.5cm\eightpoint{\bf Figure 1.}
Positive-energy part of the Hofstadter butterfly.}
\leftline{\hskip 3.0cm\eightpoint
The largest regions are for $k=1$ (left) and $k=-1$ (right).}
\vskip 0.4cm

\smallskip
A solution $u$ of the equation \equ(SchroedingerCocycle)
defines an orbit $n\mapsto(x_n,(u_n,u_{n-1}))$
for the following map $G$:
$$
G(x,y)=(x+\alpha,A(x)y)\,,\qquad x\in X\,,\quad y\in\real^2\,.
\equation(SkewG)
$$
Here, $X$ denotes the real line $\real$ or the circle $\torus=\real/\integer$,
depending on the situation being considered.
A map of this type will be referred to as a skew-product map
over a translation of $X$, or a skew-product (map) for short.
Given this connection with dynamics,
the Hofstadter butterfly can be viewed as a two-dimensional analogue
of the Arnold tongues, which characterize resonances in circle maps.
In particular, it exhibits interesting self-similarity properties [\rRuPi,\rSatij].
This strongly suggests the use of renormalization techniques.

\smallskip
Renormalization group ({\ninerm RG}) transformations
for maps that involve irrational rotations
have been studied for a variety of systems,
from circle maps and area-preserving maps of the plane,
to skew-products of the type \equ(SkewG).
Among the many references that could be listed here are
[\rKada,\rMcK,\rMR,\rStirn,\rAvKr,\rAK].
In essence, these \RG transformations lift the Gauss map
(defined on $[0,1]$,
mapping $\alpha>0$ to the fractional part of $1/\alpha$,
and zero to zero)
to a space of dynamical systems.
In order to allow for scaling,
they are usually formulated for pairs of commuting maps.

In this paper,
we focus on the inverse golden mean $\alpha_\ast=(\sqrt{5}-1)/2$,
which is a fixed point of the Gauss map.
This allows us to consider a single \RG transformation $\buR$.
Possible applications include a description of the generalized
eigenfunction of the self-dual \AM Hamiltonian $H^{\alpha_\ast}$
for the largest energy value $E_\ast$ in its spectrum.
Another possible application concerns the self-similarity
and scaling property of the Hofstadter butterfly,
as $\alpha$ approaches $\alpha_\ast$ and $E$ approaches $E_\ast$.
This self-similarity is depicted in Figure 2.
It shows $4$ successive enlargements of the Hofstadter butterfly,
zooming in on the point $(\alpha_\ast,E_\ast)$.
The largest spectrum-free region in the $n$-th magnification
corresponds to a gap index $k_n=(-1)^nf(n+1)$,
where $f(m)$ denotes the $m$-th Fibonacci number.

\vskip 0.6cm
\hbox{\hskip0.5cm
\includegraphics[height=4cm,width=6cm]{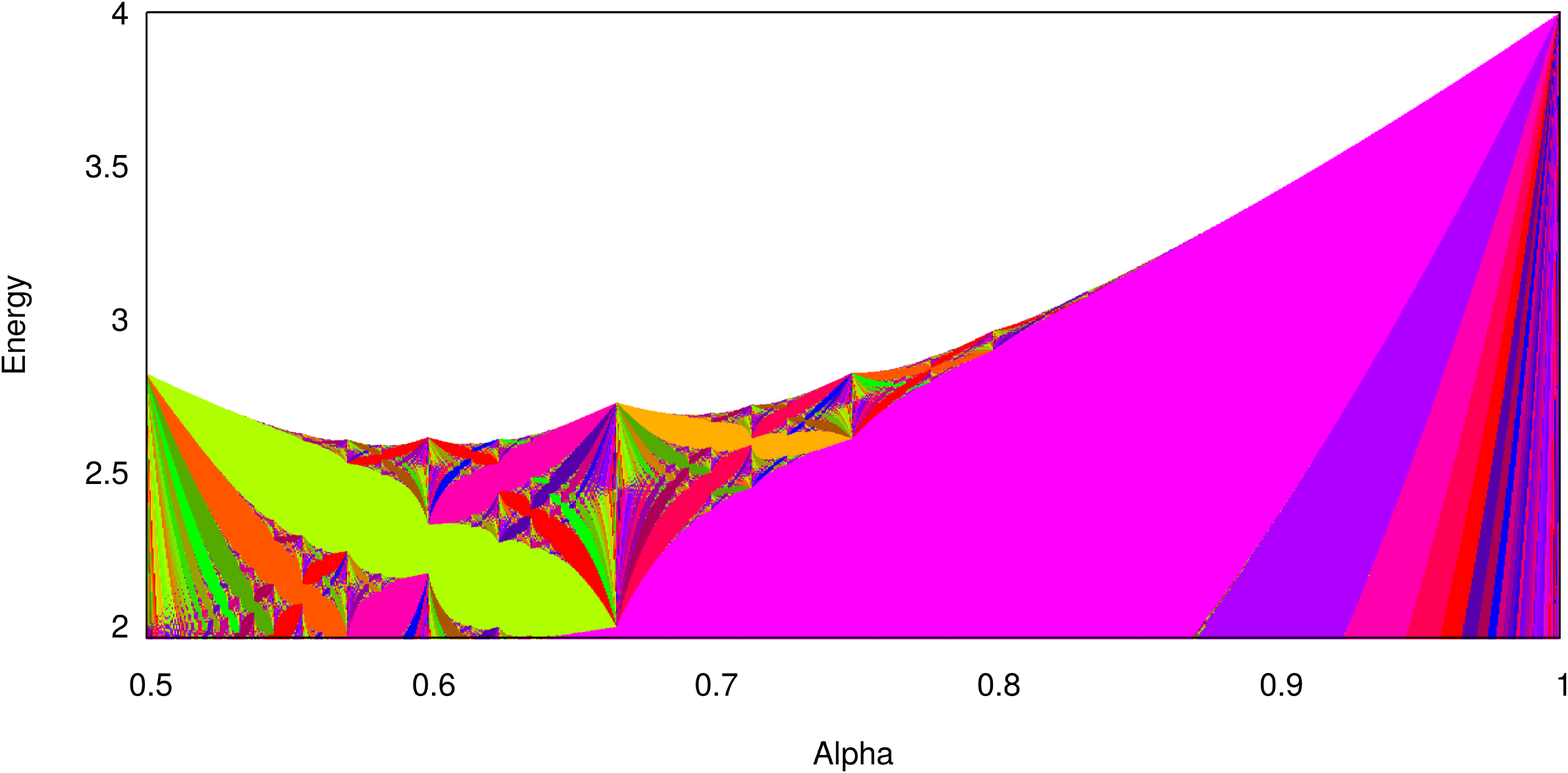}
\includegraphics[height=4cm,width=6cm]{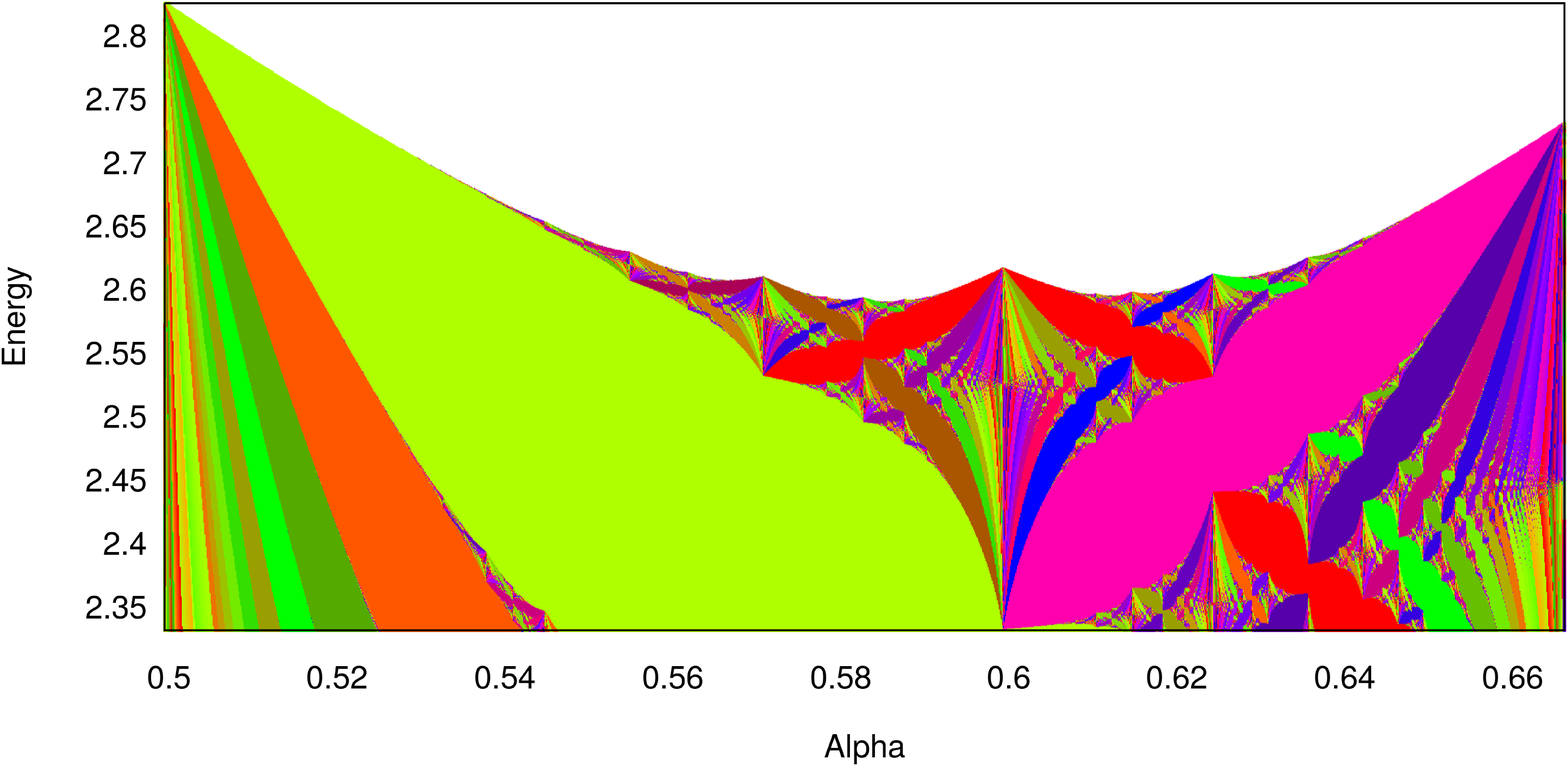}
}
\hbox{\hskip0.5cm
\includegraphics[height=4cm,width=6cm]{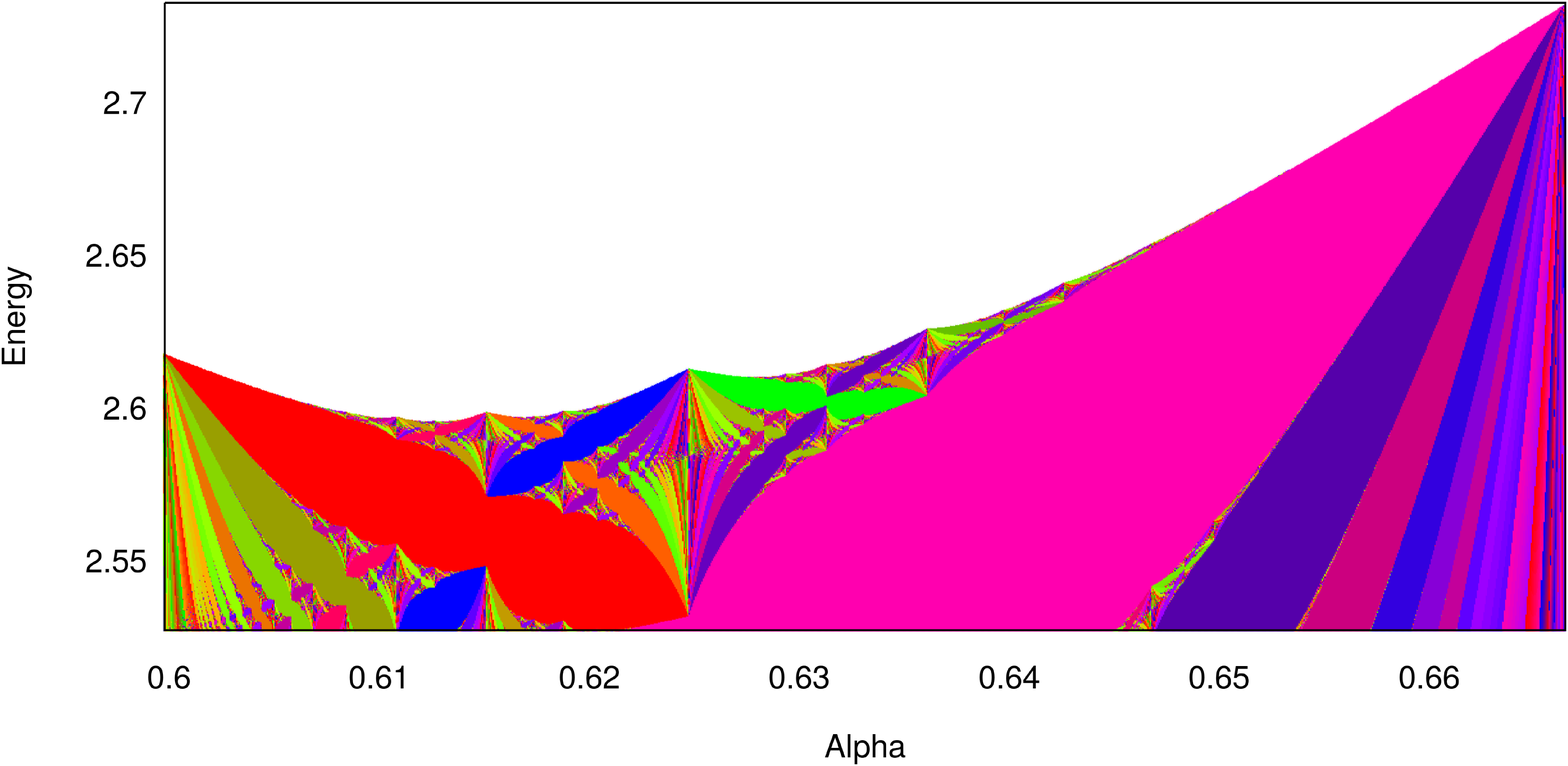}
\includegraphics[height=4cm,width=6cm]{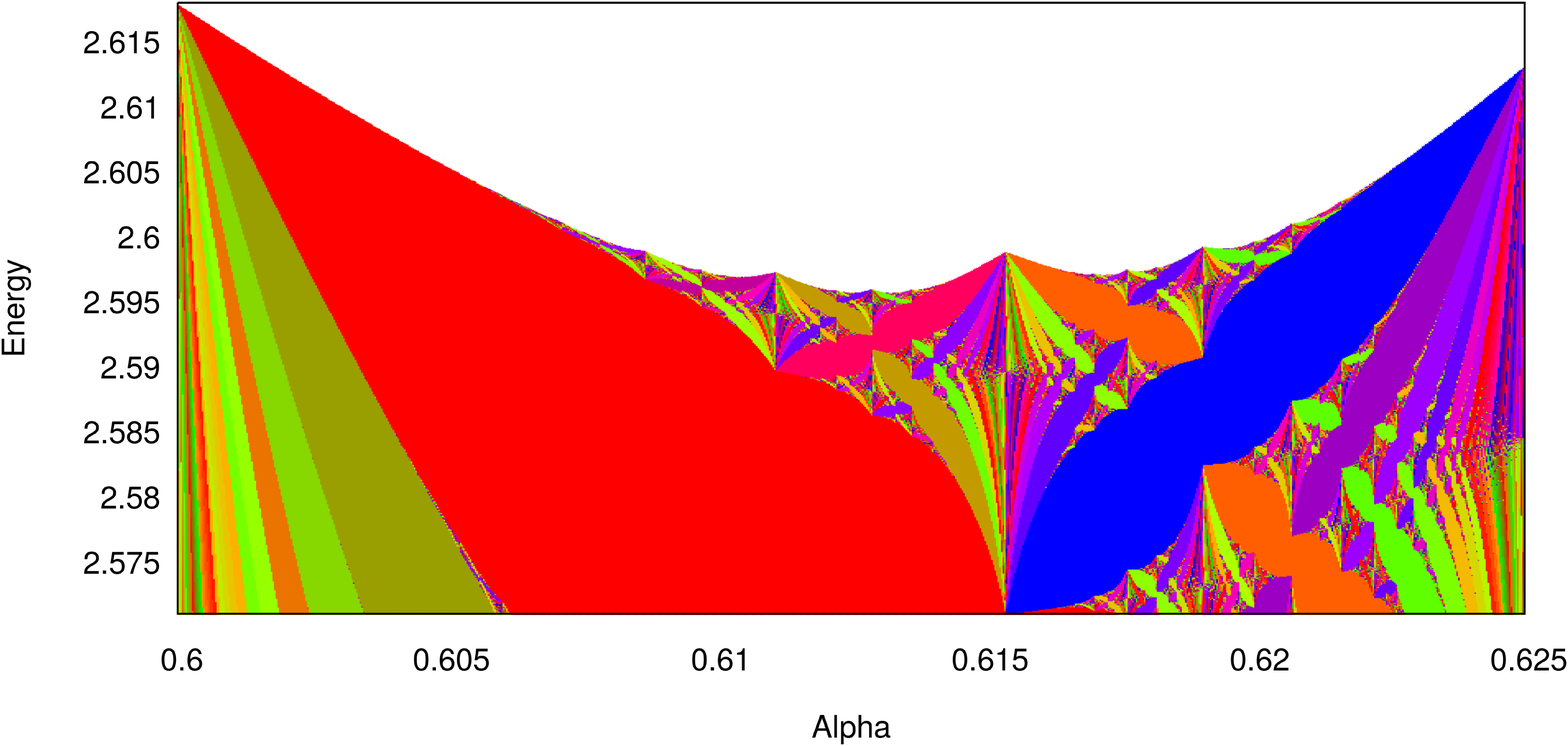}
}
\vskip 0.0cm
\leftline{\hskip 1.5cm\eightpoint{\bf Figure 2.}
Enlargements of the Hofstadter butterfly for $\alpha$ near $\alpha_\ast$.}
\vskip 0.6cm

\smallskip
In order to simplify notation, a skew-product map $G$ of the form \equ(SkewG)
will be written as $(\alpha,A)$.
Given a second map $F=(\beta,B)$ of the same type,
we define the renormalized pair as
$$
\buR(P)=\bigl(\,\Lambda_1^{-1}G\Lambda_1\,\bcomma\,\Lambda_1^{-1}FG^{-1}\Lambda_1\,\bigr)\,,
\qquad P=(F,G)\,.
\equation(singleRG)
$$
Here, $\Lambda_1$ is a map on $\real\times\real^2$
of the form $\Lambda_1(x,y)=\bigl(\alpha_\ast x,L(x)y\bigr)$,
where $L$ depends on the pair $P$ as described below.

The scaling $x\mapsto\alpha_\ast x$ of the first component is canonical and standard.
In order to motivate our choice of $L$, let us consider
the \AM map  $G=(\alpha,A)$, with $A$ given by \equ(SchroedingerCocycle)
and $V(x)=2\lambda\cos(2\pi(x+\xi))$.
Since $A$ is periodic with period $1$, $G$ commutes with $F=(1,\idmat)$,
where $\idmat$ denotes the $2\times 2$ identity map.
This property is preserved under renormalization:
if $P$ is a commuting pair, then so is $\buR(P)$.
Another noteworthy property of the transformation $\buR$ is that
it commutes with the inversion
$(F,G)\mapsto\bigl(F^{-1},G^{-1}\bigr)$ for commuting pairs,
modulo a trivial conjugacy.
This property has the potential of producing non-uniqueness,
in the sense that every \RG orbit comes in pairs.
There should be no real distinction between such orbits.
This brings us to an interesting property of the \AM map $G$:
it is reversible, in the sense that
$$
G^{-1}=\SS_c G\SS_c\,,\qquad
\SS_c(x,y)=(c-x,Sy)\,,\qquad
S=\stwomat{0}{1}{1}{0}\,,
\equation(Reversibility)
$$
with $c=\alpha-2\xi$.
Notice that $\SS_c$ is an involution, meaning that $\SS_c^2=\id$.

One of the lesson learned from the \RG analysis of area-preserving maps
[\rStirn,\rAK] is that reversibility should be preserved
under renormalization, if possible.
Thus, we choose $\Lambda_1$ to commute with $\SS_c$.
For simplicity, we consider $c=0$ and set
$$
\Lambda_1(x,y)=\bigl(\alpha_\ast x,Se^{\sigma_1S}y\bigr)\,.
\equation(SingleScaling)
$$
The constant $\sigma_1=\sigma_1(P)$ is chosen
in such a way that the renormalized pair
$\buR(P)$ satisfies a suitable normalization condition (defined later).
Unless specified otherwise,
we assume now that $\alpha=\alpha_\ast$ and $\beta=1$.
This pair of translations reproduces under renormalization,
in the sense that
$\buR(P)=\bigl(\bigl(1,B_1\bigr)\bcomma\bigl(\alpha_\ast,A_1\bigr)\bigr)$
for two matrix functions $A_1$ and $B_1$.

\smallskip
We remark that $F(x,y)$ and $G(x,y)$ need not be defined for all $x\in\real$.
Formally, if $F$ and $G$ commute,
then we can identify $F(x,y)$ with $(x,y)$
and consider $G$ to be a map on the resulting quotient space.
In any case, as far as renormalization is concerned,
it suffices that the domains of $A_1$ and $B_1$
include the domains of $A$ and $B$, respectively.

Our main result is the following.

\claim Theorem(FixedPoint)
There exists a function $A_\ast$
that is analytic on the complex disk $\bigl|x-{\alpha_\ast\over 2}\bigr|<2$,
and a function $B_\ast$ that is analytic on $\bigl|x-\half\bigr|<3$,
both non-constant and taking values in $\rmSL(2,\real)$
for real arguments, such that the following holds.
The skew-product maps $F_\ast=(1,B_\ast)$ and $G_\ast=(\alpha_\ast,A_\ast)$
are reversible and commute with each other.
Furthermore, the pair $P_\ast=(F_\ast,G_\ast)$
is a fixed point of the transformation $\buR^3$,
and the three-step scaling factor (defined later)
is given by
$$
e^{\sigma_\ast}=
1.7000157758867897671921936150581734037633645686725\ldots
\equation(expsigmaast)
$$

To our knowledge, the existence of such a $3$-periodic \RG orbit
has not been described before in the literature.
Some numerical and approximate \RG computations
can be found in [\rOsKi,\rKeSa,\rMeOsWi,\rMeOs], to mention just a few.

It is possible that the transformation $\buR$ has
other nontrivial periodic orbits, including one for zero energy.
We have not looked at this question yet\footnote{$^\dagger$}
{\eightpoint Update: In recent numerical experiments [\rKoKo]
we find a periodic orbit of lenght $6$
that attracts the self-dual {\sevenrm AM} map with zero energy.}.
The most prominent accumulation phenomenon in the Hofstadter butterfly
occurs at the point $(\alpha,E)=(0,0)$.
But this may not be within the scope of renormalization,
since the accumulation is linear, not geometric.
A scaling conjecture and some related work
can be found in [\rThou,\rLaWi,\rLasti].

\smallskip
Our proof of \clm(FixedPoint) relies on estimates
that have been carried out with the aid of a computer;
see Sections 3, 4, and 6.
As a by-product, we obtain highly accurate estimates
on various relevant quantities, including the function $A_\ast$ and $B_\ast$.
Some bounds are given in Lemma 3.1.
To be more precise about the scaling factor \equ(expsigmaast),
we note that $\buR^3$ is given by
$$
\buR^3(P)=\bigl(\,\Lambda_3^{-1}G^2F^{-1}\Lambda_3\,\bcomma\,
\Lambda_3^{-1}FG^{-1}FG^{-2}\Lambda_3\,\bigr)\,.
\equation(TripleRG)
$$
Here, $\Lambda_3$ is a composition of three scalings \equ(SingleScaling)
and thus of the form
$$
\Lambda_3(x,y)=\bigl(\alpha_\ast^3 x,Se^{\sigma_3 S}y\bigr)\,.
\equation(TripleScaling)
$$
The scaling parameter $\sigma_3=\sigma_3(P)$ is determined by a suitable
normalization condition for the pair $\buR^3(P)$.
For the precise definition we refer to Section 3.
The constant $\sigma_\ast$ that appears in \equ(expsigmaast)
is the value of $\sigma_3(P_\ast)$.
It is independent of the choice of normalization.

Following an idea that was used in [\rStirn,\rAK],
we solve the fixed point equation for $\buR^3$
by first solving the fixed point equation for
the following ``palindromic'' modification:
$$
\buR_3(P)=\bigl(\,\Lambda_3^{-1}GF^{-1}G\Lambda_3\,\bcomma\,
\Lambda_3^{-1}G^{-1}FG^{-1}FG^{-1}\Lambda_3\,\bigr)\,.
\equation(PalindromicRG)
$$
Clearly, $\buR_3(P)$ agrees with $\buR^3(P)$, if $P$ is a commuting pair.
The advantage of the transformation $\buR_3$ is that
it preserves reversibility, even for pairs that do not commute.
The condition $FG=GF$ is very inconvenient to work with,
so we drop it while solving the fixed point equation for $\buR_3$.
Once a solution $P_\ast$ is found,
it is not too hard to show that $F_\ast$ and $G_\ast$ have to commute.

\smallskip
At this time, our evidence that the behavior of $\buR$ near $P_\ast$
describes properties of the spectrum and generalized eigenfunctions
for the self-dual \AM model is purely numerical.
Our numerical results are described in Section 2.
In particular, they indicate that the following
applies to the self-dual \AM model with $\alpha=\alpha_\ast$ and $E=E_\ast$.

\claim Theorem(StableSpec)
Let $G=(\alpha_\ast,A)$ be a continuous skew-product map on $\torus\times\real^2$,
such that $P=((1,\idmat),G)$ is infinitely renormalizable.
To be more precise, write $P_n=\buR^n(P)$
as $P_n=((1,B_n)\,\bcomma\,(\alpha_\ast,A_n))$.
Assume that the sequence $n\mapsto A_n(x)$ is bounded for some $x$,
and that $\sigma_3(P_{3k})>0$ for large $k$.
Then $G$ has a nontrivial orbit that returns infinitely often
to some fixed bounded set.
In particular, if $A$ is of the form {\rm\equ(SchroedingerCocycle)},
then $E$ belongs to the spectrum
of the corresponding Schr\"odinger operator {\rm\equ(SchroedingerOp)}.

A proof of \clm(StableSpec) is given in Section 5.

We note that the asymptotic condition $\sigma_3(P_{3k})>0$
holds e.g.~if $B_{3k}\to B_\ast$ and $A_{3k}\to A_\ast$,
uniformly on the interval $(-2,2)$.

\section Some numerical results and observations

Figure 3 shows the matrix $A_\ast$ described in \clm(FixedPoint)
as a function of $x$.
To be more specific, let us first change basis and write
$A_\new=MA_\old M$ and $S_\new=MS_\old M$, with $M=M^{-1}$ as defined below.
The matrices $A=A_\new$ and $S=S_\new$ are of the form
$$
A=\twomat{t+s}{u}{v}{t-s}\,,\qquad
S=\twomat{1}{0}{0}{-1}\,,\qquad
M={1\over\sqrt{2}}\twomat{1}{1}{1}{-1}\,.
\equation(AnewSnewM)
$$
{}From now on, reversibility is defined
with respect to this new matrix $S$.
Notice that the matrix part of the scaling $\Lambda_3$
is diagonal in these new coordinates,
with eigenvalue entries $e^{\sigma_3 S}$ and $-e^{-\sigma_3 S}$.

In this representation, the Schr\"odinger matrix \equ(SchroedingerCocycle)
corresponds to $t=(E-V)/2$, $u=t+1$, $v=t-1$, and $s=0$.
If $A$ is the second component of a map $G=(\alpha,A)$,
then we usually work with the translated matrix $A_0(x)=A\bigl(x-{\alpha\over 2}\bigr)$,
so that $G$ is reversible if and only if the components
$t_0$, $u_0$, $v_0$ of $A_0$ are even, and $s_0$ is odd.
These components for the matrix $A_\ast$ are shown in Figure 3.
Judging from a few thousand Taylor coefficients,
these functions have much larger domains
than those described in \clm(FixedPoint), and we suspect
that $A_\ast$ and $B_\ast$ are in fact entire analytic.

\vskip 0.6cm
\hbox{\hskip0.2cm
\includegraphics[height=5cm,width=13cm]{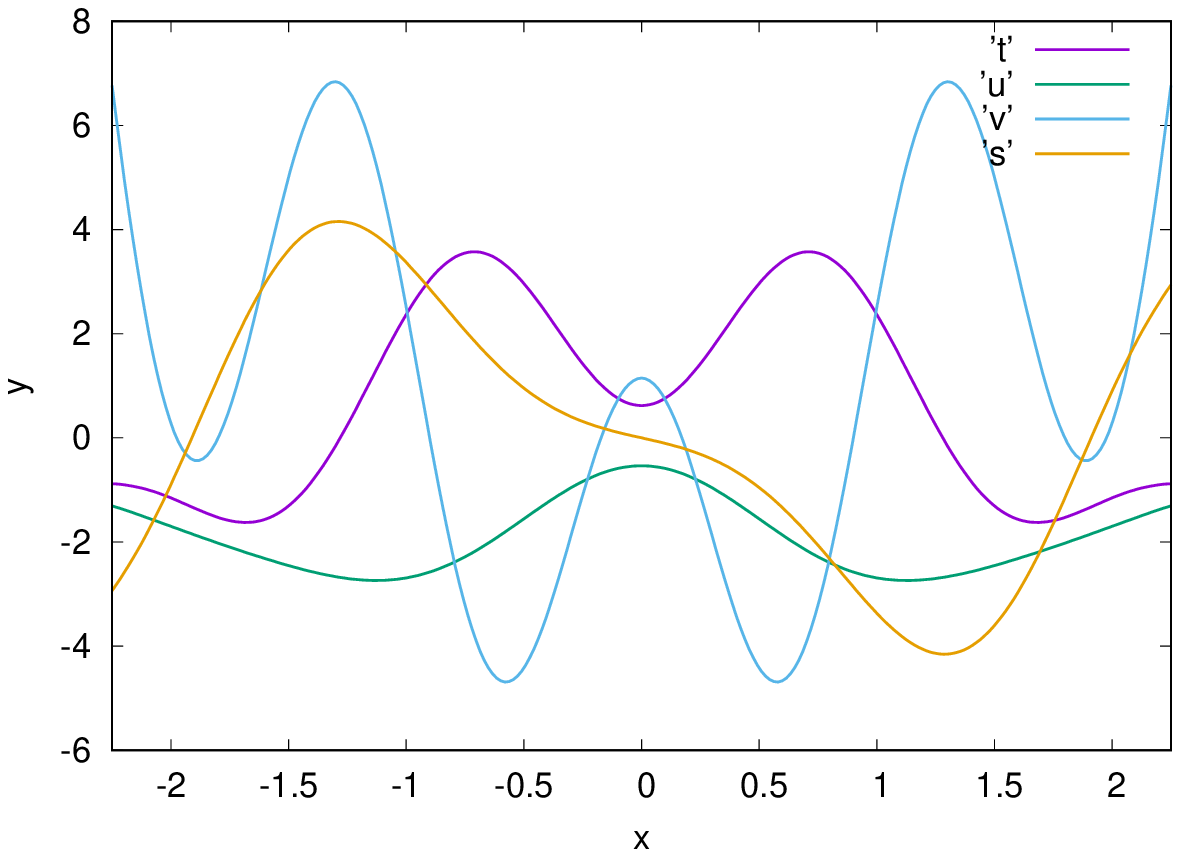}
}
\vskip 0.0cm
\leftline{\hskip 1.5cm\eightpoint{\bf Figure 3.}
Components of the matrix $A_0$ for the skew-product map $G_\ast$.}
\vskip 0.4cm

\smallskip
Our proof of \clm(FixedPoint) involves the use of an approximate
fixed point $\bar P$ for $\buR_3$.
A first rough approximation was found by computing
iterates $P_n=\buR^n(P)$ for the self-dual \AM model with $\alpha=\alpha_\ast$,
while adjusting the energy (via bisection)
to get $k\mapsto P_{3k}$ to converge numerically.
Better approximations are then obtained easily
by using the contraction $\buM$ described in Section 3.

The approximate eigenfunction $u$ mentioned in \clm(StableSpec)
is shown in Figure 4, for the self-dual \AM map
with $\alpha=\alpha_\ast$,
energy $E_\ast=2.5975151853767716484693511092199\ldots$,
and starting point $x_0+\xi=\alpha/2$.
The vector $y_n=(u_{n-1},u_n)$ for $n=0$ is the expanding eigenvector
$\bigl[{1\atop 0}\bigr]$ of the scaling $\Lambda_3$.
The vector $y_n$ at the $m$-th Fibonacci number $n=f(m)$ is again
asymptotically (for large $m$) parallel to $\bigl[{1\atop 0}\bigr]$,
with length of order $1$.

Figure 4 consists essentially of sharp peaks,
even in the ``solid'' looking regions.
The peaks that are higher than all preceding ones are at
$n=1$, $6$, $27$, $116$, $493$, $2090$, $8855$, $37512$,
$158905$, $673134$, $2851443$, $\ldots$
These values $n(1)$, $n(2)$, $\ldots$ fit the formula
$$
n(m)=\thalf\bigl[f(3m+1)-1\bigr]\,.
\equation(OEISseq)
$$
The \RG period $3$ is clearly visible in these data.
Notice that $n(m)\sim\alpha_\ast^{-3m}$,
and the corresponding peaks in Figure 4 grow like $e^{2m\sigma_\ast}$.
The sequence \equ(OEISseq) appears in other contexts as well
and is listed as A049651 in the On-Line Encyclopedia of Integer Sequences.
References and links can be found at [\rOEIS].

Another property of the orbit $u$ depicted in Figure 4
is that $u_n\ge 0$ for all $n$.
This indicates that the \AM map $G$ for $E=E_\ast$
has a zero rotation number.
For values of $E$ below $E_\ast$, we find positive rotation numbers.

\vskip 0.6cm
\hbox{\hskip-0.2cm
\includegraphics[height=5cm,width=13.5cm]{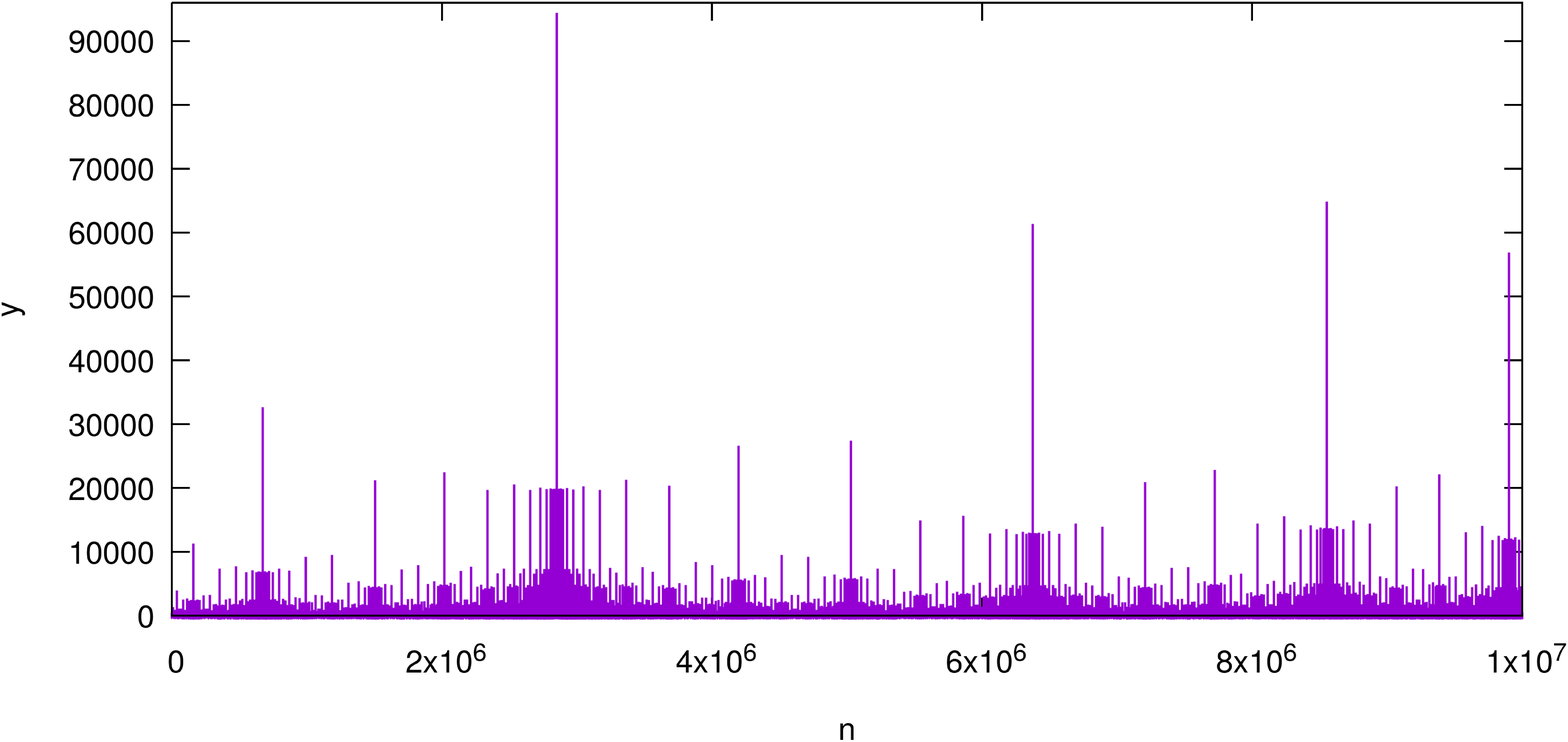}
}
\vskip0.2cm
\leftline{\hskip-0.2cm\eightpoint{\bf Figure 4.}
Generalized eigenfunction for the self-dual AM Hamiltonian
with $\alpha=\alpha_\ast$ and $E=E_\ast$.}
\vskip 0.4cm

Our proof of \clm(FixedPoint) also involves the use of
of a modest-size matrix approximation for the derivative $D\buR_3(P_\ast)$.
By increasing the dimension to get a more accurate approximation,
the eigenvalues of modulus larger than $1/10$ are found to be
$$
\eqalign{
\mu_1&=\,30.79005494022096246\ldots\,,\hskip 20pt
\mu_{3,4}=\pm 0.68224911725088276\ldots\,,\cr
\mu_2&=\nominus 4.23606797749978969\ldots\,,\hskip 26pt
\mu_5= -0.13757909772243458\ldots\,.\cr}
\equation(topEigen)
$$
The largest eigenvalue, $\mu_1$, is almost certainly
related to the (three generation)
scaling of the Hofstadter butterfly in the energy direction.
The scaling seen in Figures 1 and 2, averaged
over $4$ generations, agrees quite well with $\mu_1^{1/3}$.
The scaling in the $\alpha$-direction
over $3$ generation is trivially $-\alpha_\ast^{-6}$.
But our current \RG analysis is for fixed $\alpha=\alpha_\ast$,
so there is no room for an eigenvector of $D\buR_3(P_\ast)$
in the direction of a change of $\alpha$.

Concerning the eigenvalue $\mu_2$,
we conjecture that its value is equal to $\alpha_\ast^{-3}$.
But despite its ``trivial'' value, it is not associated
with a coordinate change or a non-commuting direction.
We believe that $\mu_2$ is related to variations in
the strength of the $x$-dependence.
In the \AM model, such a change characterizes the
transition between the conducting phase $\lambda<1$
and the insulating phase $\lambda>1$.
So far, we have not found a way to prove
that this eigenvalue is indeed $\alpha_\ast^{-3}$.
But some formal arguments are given in Section 6.

\smallskip
The eigenvalue $\mu_4$
is most likely associated with a coordinate change
and has the value $\mu_4=-e^{2\sigma_\ast}\alpha_\ast^3$.
Our program finds an additional eigenvalue $-1$
that we have omitted from the list \equ(topEigen).
We believe that this eigenvalue is associated with a non-commuting direction,
which makes it irrelevant for commuting pairs of maps.

A more curious observation is that many (if not all)
contracting eigenvalues other than $\mu_5$
appear in pairs of opposite sign.
This is not unusual for some ``trivial'' eigenvalues,
as will be described in Section 6,
but we have no explanation why the same might occur more generally.

\section The fixed point problem

In this section we reformulate the equation $\buR_3(P)=P$
as a fixed point problem for a contraction,
acting on a suitable space of pairs.

\subsection Normalization

Since the transformation $\buR_3$ involves
the composition and inverses of skew-product maps,
let us first describe these two operations.
As mentioned in the last section,
the matrix part $A$ of a map $G=(\alpha,A)$ is being represented
as $A=A_0\bigl({\alpha\over 2}+\bdot\bigr)$.
Then $G$ is reversible if and only if $A_0^{-1}(x)=SA_0(-x)S$ for all $x$.
The composition of two skew-products is given by
$$
\textstyle
(\beta,B)(\alpha,A)=(\alpha+\beta,C)\,,\qquad
C_0=B_0\bigl({\alpha\over 2}+\bdot\bigr)
A_0\bigl(-{\beta\over 2}+\bdot\bigr)\,.
\equation(oRepProdMat)
$$
In particular, $(\beta,B)$ is the inverse of $(\alpha,A)$
if and only if $\beta=-\alpha$ and $B_0=A_0^{-1}$.

\smallskip
Consider now a conjugacy $H\mapsto\Lambda_3^{-1}H\Lambda_3$.
In the expression \equ(PalindromicRG) for $\buR_3(P)$,
such a conjugacy is being applied to $GF^{-1}G$ and $G^{-1}FG^{-1}FG^{-1}$.
Consider first $H=GF^{-1}G$,
which is of the form $H=(\alpha_\ast^3,C)$.
The matrix part of $\Lambda_3^{-1}H\Lambda_3$ is given by
$$
e^{-\sigma_3S}SCSe^{\sigma_3S}
=\twomat{t+s}{-e^{-2\sigma_3}u}{-e^{2\sigma_3}v}{t-s}\,,\quad{\rm if}\quad
C=\twomat{t+s}{u}{v}{t-s}\,.
\equation(Equalize)
$$
Our normalization condition that determines $\sigma_3$
is that $e^{-2\sigma_3}u_0(0)$ and $e^{2\sigma_3}v_0(0)$
have the same absolute value.
Clearly, other normalization conditions would work equally well.
The same value of $\sigma_3$ is used to scale $H=G^{-1}FG^{-1}FG^{-1}$.
In other words, only the first component of the
pair $\buR_3(F,G)$ is being ``re-normalized''.
But of course, this affects both components
when $\buR_3$ is being iterated.

\subsection An extension

Given the constructive nature of our analysis,
an important question is how to deal with a constraint like $\det(A)=1$.
Typical $\rmSL(2,\real)$ methods,
including an Iwasawa-type decomposition for real matrices,
involve quantities that have singularities in the complex plane.
The resulting bounds were not sufficient for our purpose.
For the problem considered here, it is better to consider
$\rmPSL(2,\complex)$, via M\"obius transformations
$$
\bma z={az+u\over vz+d}\,,\qquad
A=\twomat{a}{u}{v}{d}\,.
\equation(MoebiusA)
$$
In particular, our involution is represented by
$$
\bms z=-z\,,\qquad
S=\stwomat{1}{0}{0}{-1}\,.
\equation(MoebiusS)
$$
Notice that the transformation $\bma$ is well-defined as long as $ad-uv\ne 0$.
Our maps $G=(\alpha,A)$ involve matrices $A\in\rmSL(2,\real)$,
so the corresponding M\"obius transformations $\bma$
map the upper half of the extended complex plane $\complex\cup\{\infty\}$
into itself.
As long as $ad-uv\ne 0$, we have
$$
\bma^{-1}z={dz-u\over-vz+a}\,,\qquad
\bms\bma\bms z={az-u\over-vz+d}\,.
\equation(MoebiusSAS)
$$
Consider temporarily $G=(\alpha,\bma)$ instead of $G=(\alpha,A)$.
We say that $G$ is reversible if $\SS_0G\SS_0=G^{-1}$,
where $\SS_0(x,z)=(-x,\bms z)$.
For the translated quantities described
after \equ(AnewSnewM) and at the beginning of this section,
reversibility means that
$$
\bma_0(x)^{-1}=\bms\bma_0(-x)\bms\,,\qquad
{d_0(x)z-u_0(x)\over-v_0(x)z+a_0(x)}
={a_0(-x)z-u_0(-x)\over-v_0(-x)z+d_0(-x)}\,.
\equation(MoebiusRev)
$$
In other words, the functions $t_0=(a_0+b_0)/2$, $u_0$, $v_0$ are even,
and $s_0=(a_0-b_0)/2$ is odd.
Notice that this does not require that $A_0$ has determinant $1$.
And the same applies to the expression $\bigl[{d~-b\atop-c~a}\bigr]$
for the matrix representing the inverse $\bma^{-1}$.

So for all practical purposes, the constraint $\det(A)=1$ has been eliminated,
albeit at the cost of having more degrees of freedom than necessary.

\smallskip
Motivated by the above,
we extend our \RG transformation $\buR_3$ to pairs of maps $P=(F,G)$
that need that need not be area-preserving.
(We call $(\alpha,A)$ area-preserving if $A$ has determinant $1$.)
Still, it is preferable for the fixed point of $\buR_3$
to be are-preserving.
This can be done e.g.~by composing $\buR_3$ with the normalization map
$$
\buN\bigl((\alpha,A)\bigr)=\bigl(\alpha,\NN(A)\bigr)\,,\qquad
\NN(A)=[\det(A)]^{-1/2}A\,.
\equation(DivSqrtDet)
$$
If the determinant of $A$ is close to $1$,
then $[\det(A)]^{-1/2}$ is well-defined, and $\NN(A)$ has determinant $1$.
Notice also that, if $(\alpha,A)$ is reversible,
then $\det(A_0)$ is an even function, so $(\alpha,\NN(A))$ is still reversible.
The derivative of $\NN$ at $A$ is given by
$$
D\NN(A)\dot A=
\det(A)^{-1/2}\dot A
-\thalf\det(A)^{-3/2}\bigl[a\dot d+d\dot a-u\dot v-v\dot u\,\bigr]A\,.
\equation(DerNN)
$$
Our extension of $\buR_3$ is now defined as
$$
\buF=\buN\circ\buR_3\,,\qquad
\buN\bigl((F,G)\bigr)=\bigl(\buN(F),\buN(G)\bigr)\,.
\equation(ExtRGThree)
$$

We consider this map $\buF$ in a neighborhood
of an approximate fixed point $\bar P$.
In what follows, the domain of $\buR_3$ is restricted to
pairs $P=(F,G)$ whose components $F=(\beta,B)$ and $G=(\alpha,A)$
are reversible, with $\beta=1$ and $\alpha=\alpha_\ast$.
The maps $F$ and $G$ need not be area-preserving. But by construction,
$\buF(P)$ is a pair of reversible area-preserving maps.

\subsection The contraction

As is common in many computer-assisted proofs,
we convert the fixed point problem for the given map $\buF$
to a fixed point problem for a quasi-Newton map
$\buM$ associated with $\buF$.
To be more specific, let $\id-M$ be an approximate inverse
of $\id-D\buF\bigl(\bar P\bigr)$.
Then we define
$$
\buM(p)=\buF\bigl(\bar P+(\id-M)p\bigr)-\bar P+Mp\,.
\equation(ContrRGThree)
$$
Here, the sum of map-pairs is defined component-wise,
and $(\alpha,A_1)+(\alpha,A_2)$ is defined as $(\alpha,A_1+A_2)$.
If $\bar P$ is close to being a fixed point of $\buF$,
and if $M$ is chosen properly, we can expect
$\buM$ to be a contraction in some neighborhood of $\bar P$.
Notice that, if $p$ is a fixed point of $\buM$,
then $P=\bar P+(\id-M)p$ is a fixed point of $\buF$.

\smallskip
Now we need to define some function spaces.
Given $\rho>0$,
denote by $\AA_\rho$ the space of all real analytic functions $f$
on $(-\rho,\rho)$ that have a finite norm
$$
\|f\|_\rho=\sum_{n=0}^\infty|f_n|\rho^n\,,\qquad
f(x)=\sum_{n=0}^\infty f_n x^n\,.
\equation(AANorm)
$$
Of course, every $f\in\AA_\rho$
extends to an analytic function on the complex disk $|x|<\rho$.
Furthermore, $\AA_\rho$ is a Banach algebra
under the pointwise product of functions.

The space of matrix functions $A_0$
whose components $t_0$, $u_0$, $v_0$, and $s_0$
belong to $\AA_\rho$ will be denoted by $\AA_\rho^4$.
The norm of $A_0\in\AA_\rho^4$ is defined as
$\|A_0\|_\rho=\|t_0\|_\rho+\|u_0\|_\rho+\|v_0\|_\rho+\|s_0\|_\rho$.
To define a space for pairs of such functions,
we first fix a pair $\rho=(\rho_\srmF,\rho_\srmG)$
of positive real number.
Then we define $\BB_\rho$ to be the vector space of all pairs
$p=(B_0,A_0)$ in $\AA_{\rho_\ssrmF}^4\times\AA_{\rho_\ssrmG}^4$,
equipped with the norm
$\|p\|_\rho=\|B_0\|_{\rho_\ssrmF}+\|A_0\|_{\rho_\ssrmG}$.
The subspace of reversible pairs is denoted by $\BB_\rho^{\tinyskip r}$.

\smallskip
Due to the above-mentioned restrictions on the domain of $\buF$,
any skew-product $H=(\gamma,C)$ that appears
at some stage in the computation of $\buF$ or $\buM$
has a pre-determined first component $\gamma$.
Thus, in order to simplify notation related to domains and function spaces,
let us now identify such a map $H$
with its translated matrix component $C_0=C\bigl(\bdot-{\gamma\over 2}\bigr)$.

In order for $\buR$ to be defined as a map on $\BB_\rho$,
it is necessary and sufficient that
$$
\thalf\alpha^{-2}\le\rho_\srmG\,,\qquad
\thalf\alpha+\alpha\rho_\srmG\le\rho_\srmF\le\alpha^{-1}\rho_\srmG\,.
\qquad({\rm for~}\buR)
\equation(RROneDomainCond)
$$
These inequalities are easily satisfied
e.g.~with $2=\rho_\srmG\le\rho_\srmF\le 3$.
But it should be noted that, if $P$ belongs to $\BB_\rho$
with $\rho$ satisfying \equ(RROneDomainCond),
then the components of $\buR(P)$ are defined on significantly larger domains.
Those larger domains are not disks;
however, they improve the domain of iterates of $\buR$.
If we restrict to $\rho_\srmG\le\rho_\srmF$,
then the analogue of the condition \equ(RROneDomainCond)
for the transformation $\buR_3$ is
$$
\thalf\le\rho_\srmG\le
\rho_\srmF\le\alpha^{-3}\rho_\srmG-\thalf\alpha^{-1}\,.
\quad\qquad({\rm for~}\buR_3)
\equation(RRThreeDomainCond)
$$
This condition is significantly weaker than \equ(RROneDomainCond).

\claim Lemma(Contr)
Let $\rho=(3,2)$.
Then there exist a pair $\bar P$ in $\BB_\rho^{\tinyskip r}$,
a bounded linear operator $M$ on $\BB_\rho^{\tinyskip r}$,
and positive constants $\eps,K,\delta$ satisfying
$\eps+K\delta<\delta$, such that
the transformation $\buM$ defined by \equ(ContrRGThree)
is analytic in $B_\delta$ and satisfies
$$
\|\buM(0)\|_\rho\le\eps\,,\qquad
\|D\buM(p)\|_\rho\le K\,,\qquad p\in B_\delta\,,
\equation(ContrBounds)
$$
where $B_\delta$ denotes the open ball of radius $\delta$
in $\BB_\rho^{\tinyskip r}$, centered at the origin.
Furthermore, for every pair $p\in B_\delta$,
the matrix components of $P=\bar P+(\id-M)p$ are non-constant,
$e^{\sigma_3(P)}$ satisfies the bound
defined by the right hand side of \equ(expsigmaast),
and $\bigl\|P-\bar P\bigr\|_\rho<10^{-280}$.

Our proof of \clm(Contr) is computer-assisted
and will be described in Section 7.
We note that much higher precisions than the one described
in this lemma can be achieved quite easily.

\section Proof of \clm(FixedPoint)

Assume that \clm(Contr) holds.
By the contraction mapping principle, the given bounds imply
that $\buM$ has a unique fixed point $p_\ast$ in $B_\delta$.
The corresponding function $P_\ast=\bar P+(\id-M)p_\ast$
is a fixed point of $\buF$,
and the last statement in \clm(Contr) applies to $p=p_\ast$.

What remains to be proved is that the maps
$F_\ast$ and $G_\ast$ commute.
To this end,
consider the commutator $\Theta=FG(GF)^{-1}$
for a general pair $P=(F,G)$.
The commutator for the renormalized pair $\tilde P=\buR_3(P)$
is easily found to be
$$
\tilde\Theta=(G\Lambda)^{-1}\Theta^{-1}(G\Lambda)\,.
\equation(tildeTheta)
$$
If we write $\Theta=(0,C)$,
then $\tilde\Theta=\bigl(0,\tilde C\bigr)$, with
$$
\tilde C(x)=e^{-\sigma_3 S}S
A\bigl(\alpha^3x\bigr)^{-1}C\bigl(\alpha^3x+\alpha\bigr)^{-1}
A\bigl(\alpha^3x\bigr)Se^{\sigma_3 S}\,.
\equation(tildeThetamat)
$$
Consider a change of variables $x={1\over 2\alpha}+z$.
Define $C_1(z)=C(x)$ and $\tilde C_1(z)=\tilde C(x)$.
Then the equation \equ(tildeThetamat) becomes
$$
\tilde C_1(z)
=A_1(z)^{-1}C_1\bigl(\alpha_\ast^3z\bigr)^{-1}A_1(z)\,,\qquad
A_1(z)=A\bigl(\thalf\alpha_\ast^2+\alpha_\ast^3z\bigr)Se^{\sigma_\ast S}\,.
\equation(CiOne)
$$
Let now $P=P_\ast$, so that $\tilde C_1=C_1$.
We need the identity \equ(CiOne)
in some (arbitrary small) complex open neighborhood of the origin.
It is straightforward to check that all these matrix functions
are being evaluated only at points in their domain.
Taking the trace on both sides of \equ(CiOne) yields
$\tr(C_1(z))=\tr(C_1\bigl(\alpha^3z\bigr))$.
By analyticity, this implies that the trace
of $C_1(z)$ is independent of $z$,
and the same holds then for the eigenvalues.

Assume now that the following holds for our fixed point $P_\ast$.

\claim Proposition(AiCi)
The matrix $A_1(0)=A_0\bigl(\half\bigr)Se^{\sigma_\ast S}$
has no real or imaginary eigenvalues,
and the matrix $C_1(0)$ does not have an eigenvalue $-1$.

Applying \equ(CiOne) twice, we also have
$C_1(0)=A_2(0)^{-2}C_1(0)A_1(0)^2$.
In other words, $C_1(0)$ commutes with $A_1(0)^2$.
Consider now a basis in $\complex^2$ where $A_1(0)$ is diagonal.
By \clm(AiCi), such a basis exists.
Then $A_2(0)$ is diagonal as well,
and its eigenvalues are non-real by \clm(AiCi).
So the matrix $C_1(0)$ has to be diagonal as well;
and in particular, it commutes with $A_1(0)$.
Now \equ(CiOne) implies that $C_1(0)$ is its own inverse.
And $C_1(0)$ has no eigenvalue $-1$ by \clm(AiCi).
So $C_1(0)$ must be the identity matrix.
Given that $C_1(z)$ is independent of $z$,
we conclude that $\Theta=(0,\idmat)$,
or equivalently, that $F_\ast$ and $G_\ast$ commute.

This concludes the proof of \clm(FixedPoint),
conditioned on the validity of \clm(Contr) and \clm(AiCi).

\section Recurrent orbits

The main goal here is to give a proof of \clm(StableSpec).
Let $P=(F,G)$ be a commuting pair of skew-products
$F=(1,B)$ and $G=(\alpha_\ast,A)$,
where $A$ and $B$ are functions with values in $\rmSL(2,\real)$.
Assume that the renormalized maps
$$
F_n=(1,B_n)\,,\quad
G_n=(\alpha_\ast,A_n)\,,\quad
(F_n,G_n)=P_n\defeq\buR^n(P)\,,
\equation(FnGnDef)
$$
are all well-defined.
This involves a condition on the (real) domains of $A$ and $B$.
It suffices e.g.~that $F$ be defined on $I_\srmF=(-\rho_\srmF,\rho_\srmF)$
and $G$ on $I_\srmG=(-\rho_\srmG,\rho_\srmG)$,
with $\rho_\srmF$ and $\rho_\srmG$ satisfying \equ(RROneDomainCond).
But in order to avoid domain issues when re-arranging factors,
assume that $F$ and $G$ are skew-products on $\torus\times\real^2$.

Let $n\mapsto q_n$ be the Fibonacci sequence,
defined recursively via $q_0=0$, $q_1=0$,
and $q_{n+1}=q_n+p_n$ for $n\ge 1$, where $p_n=q_{n-1}$.
Given that $F$ and $G$ commute, we have
$$
\eqalign{
F_n&=\Lambda_n^{-1}F^{p_{n-1}}G^{-q_{n-1}}\Lambda_n\,,
\qquad G_n=\Lambda_n^{-1}F^{-p_n}G^{q_n}\Lambda_n\,,
\qquad (n {\rm~even})\,,\cr
F_n&=\Lambda_n^{-1}F^{-p_{n-1}}G^{q_{n-1}}\Lambda_n\,,
\qquad G_n=\Lambda_n^{-1}F^{p_n}G^{-q_n}\Lambda_n\,.
\qquad\; (n {\rm~odd})\,,\cr}
\equation(FnGn)
$$
with $\Lambda_n$ being a scaling of the form
$$
\Lambda_n(x,y)=\bigl(\alpha_\ast^n,S^n e^{\sigma_n S}\bigr)\,.
\equation(nScaling)
$$
Here, $\sigma_n$ is a sum of scaling exponents.
More specifically,
if $n$ is a multiple of $3$, say $n=3k$, then $\sigma_n$ is the sum
of all exponents $\sigma_3(P_{3m})$ with $m<k$.
If $n$ is even, then \equ(FnGn) yields
$$
F^{-p_n}G^{q_n}\bigl(\alpha_\ast^nx,y\bigr)
=\bigl(\alpha_\ast^n(x+\alpha_\ast),e^{\sigma_n S}A_n(x)e^{-\sigma_n S}y\bigr)\,.
\equation(FnegpnGqn)
$$
A similar identity is obtained if $n$ is odd.
But in order to prove \clm(StableSpec),
it suffices to consider even $n$.
Let $y=\bigl[{1\atop 0}\bigr]$, so that $Sy=y$.
Then the second component in \equ(FnegpnGqn) is given by
$$
y_n\defeq e^{\sigma_n S}A_n(x)e^{-\sigma_n S}y
=e^{\sigma_n(S-\idmat)}A_n(x)y\,.
\equation(ynDef)
$$
Assume now that the sequence $n\mapsto A_n(x)$ is bounded
for some fixed value of $x$ in the domain of the functions $A_n$.
Assume furthermore that $\sigma_n$ is positive for sufficiently large $n$.
This holds e.g.~if $B_{6k}\to B_\ast$ and $A_{6k}\to A_\ast$,
uniformly on $I_\srmF$ and $I_\srmG$, respectively,
since $\sigma_\ast$ is positive by \equ(expsigmaast).
Given that $S-\idmat\le 0$, we see from \equ(ynDef) that
the sequence $n\mapsto y_n$ is bounded.

Assume now that $F=(1,\idmat)$.
In this case, $G^{q_n}(x,y)=\bigl(\alpha_\ast^n(x+\alpha_\ast),y_n\bigr)$.
So the above implies that $G$ has an orbit that returns infinitely often
to a fixed bounded set in $\torus\times\real^2$,
as was claimed in \clm(StableSpec).
The assertion concerning Schr\"odinger operators
is an immediate consequence of this recurrence.

\section Some trivial eigenvalues

A well-known source of trivial eigenvalues
in the renormalization of dynamical systems are coordinate changes.
For pairs of maps,
another source can be the scaling behavior of the commutator;
see e.g.~[\rKoch].
For the skew-product maps considered here, there may be
another quantity whose scaling produces a trivial value $\alpha^{-3}$
for the eigenvalue $\mu_2$.
A possibility will be mentioned at the end of this section.
Since the spectrum of $D\buR_3(P_\ast)$ is not the main topic of this paper,
we shall keep this section short and thus mostly formal.

\subsection Coordinate changes

For simplicity, let us replace the scaling $\Lambda_3$ in the
definition \equ(PalindromicRG) of $\buR_3$
by the scaling $\Lambda_\ast$ for the fixed point $P_\ast$.
This produces some extra eigenvalues for $D\buR_3(P_\ast)$,
but these can easily be identified.
Under a change of coordinates $H_\eps$ we have
$$
\buR_3\bigl(H_\eps^{-1}P_\ast H_\eps\bigr)
=\bigl(\Lambda_\ast^{-1}H_\eps\Lambda_\ast\bigr)^{-1}
P_\ast\bigl(\Lambda_\ast^{-1}H_\eps\Lambda_\ast\bigr)\,.
\equation(RGinvHepsPHeps)
$$
Setting $H_\eps=\id+\eps\dot H+\OO(\eps^{\sss 2})$
and differentiating with respect to $\eps$ yields
$$
D\buR_3(P_\ast)P_{\dot H}
=P_{\Lambda_\ast^{-1}\dot H\Lambda_\ast}\,,\qquad
P_{\dot H}\defeq{d\over d\eps}H_\eps^{-1}P_\ast H_\eps\Bigm|_{\eps=0}\,,
\equation(DRPdotH)
$$
with the map $\dot H\mapsto P_{\sss\dot H}$ being linear.
In particular, if
$\Lambda_\ast^{-1}\dot H\Lambda_\ast=\kappa\dot H$,
then $P_{\dot H}$ is an eigenvector of $D\buR_3(P_\ast)$
with eigenvalue $\kappa$.

Since our analysis is for fixed circle rotations,
let us consider just $\dot H=(0,\dot C)$.
Near the origin we have $\dot C(x)=x^n[\CC_n+\oo(1)]$
for some nonnegative integer $n$.
Then the eigen-equation $\Lambda_\ast^{-1}\dot H\Lambda_\ast=\kappa_n\dot H$
yields
$$
\alpha_\ast^{3n}\SS^{-1}\CC_n\SS=\kappa_n\tinyskip\CC_n\,,\qquad
\SS=Se^{\sigma_\ast S}=\diag\bigl(e^{\sigma_\ast},-e^{-\sigma_\ast}\bigr)\,.
\equation(CdotnEigen)
$$
So either $\kappa_n=\alpha_\ast^{3n}$ and $\CC_n$ is diagonal
(we may assume that the trace is zero),
or else $\kappa_n=-e^{\pm 2\sigma_\ast}\alpha_\ast^{3n}$
and $\CC_n$ has a single nonzero entry, off the diagonal.
Many of these eigenvalues are indeed observed numerically,
but only for $n>0$.

\subsection Commutators

Let $P=(F,G)$ with $F=(1,B)$ and $G=(\alpha_\ast,A)$.
We assume that $A=A_\ast+\OO(\eps)$ and $B=B_\ast+\OO(\eps)$
depend smoothly on a parameter $\eps$.
Notice that, to first order in $\eps$,
the right hand side of \equ(tildeTheta) depends on $\eps$
only through the factor $\Theta^{-1}$.
Consider now the equation \equ(CiOne)
that relates the commutator $\bigl(0,\tilde C\bigr)$
for the renormalized pair $\tilde P=\buR_3(P)$
to the commutator $(0,C)$ for $P$.
Substituting
$C_1=\idmat+\eps\CC+\OO(\eps^{\sss 2})$
and $\tilde C_1=\idmat+\eps\tilde\CC+\OO(\eps^{\sss 2})$ into \equ(CiOne),
and equating terms of order $\eps$, we obtain
$$
\tilde\CC(z)
=-A_1(z)^{-1}\CC\bigl(\alpha_\ast^3 z\bigr)A_1(z)\,.
\equation(tildeCCfromCC)
$$
Consider now an eigenvector of $\CC\mapsto\tilde\CC$.
Near $z=0$ we have $\CC(z)=z^n[C_n+\oo(1)]$
for some nonnegative integer $n$.
Denoting the eigenvalue by $\eta_n$, we must have
$$
\eta_n\tinyskip\CC_n=-\alpha_\ast^{3n} A_1(0)^{-1}\CC_nA_1(0)\,.
\equation(CCnEigen)
$$
Recall from \clm(AiCi) that $A_1(0)$
has two distinct eigenvalues $\theta$ and $\bar\theta=\theta^{-1}$
whose squares are non-real.
This implies e.g.~that there exists a nonzero linear combination
of $\idmat$ and $A_1(0)$ that has a zero trace.
This yields a solution $\CC_n$ of \equ(CCnEigen)
with eigenvalue $\eta_n=-\alpha_\ast^{3n}$.
Many of these eigenvalues are indeed observed in our computations,
including $\eta_0=-1$.
The non-real solutions $\eta_n=-\theta^{\pm 2}\alpha^{3n}$
are not observed (within the accuracy used).
This indicates that non-commuting perturbations contract
under renormalization,
with the possible exception of one direction with eigenvalue $-1$.
We note that this applies to $\buR_3$
but not necessarily $\buR^3$.

\demo Remark(PlusMinusEigenPairs)
The equations \equ(CdotnEigen) and \equ(CCnEigen)
are merely restrictions on eigenvalues
that could arise from coordinate transformations and commutators,
respectively.
To find out more, one needs to determine the associated eigenvectors.
If an eigenvector violates a constraint like reversibility,
or if it is due to having replaced $\Lambda_3$ by $\Lambda_\ast$,
then it is not observed in our analysis.

\subsection The second largest eigenvalue

We conclude this section with two formal arguments
supporting the conjecture that the derivative of $\buR^3$
at $P_\ast$ has an eigenvalue $\alpha_\ast^{-3}$
associated with a change of the strength of the $x$-dependence.

Consider the \RG iterates $(F_n,G_n)$ for a commuting pair $(F,G)$,
as described by the equation \equ(FnGnDef).
Taking $F=(1,\idmat)$,
the matrix part $A_n$ of $G_n$ has the trace
$$
\tr\bigl(A_n(x)\bigr)
=\tr\bigl(\PP_{q_n}(\alpha,\alpha_\ast^n x)\bigr)\,,
\quad
\PP_q(\alpha,x)\defeq A(x+(q-1)\alpha)\cdots A(x+\alpha)A(x)\,.
\equation(trAnx)
$$
Here $q_n$ denotes the $n$-th Fibonacci number.
Let now $G$ be the \AM map with $\lambda\le 1$,
and with $\xi=0$ for simplicity.
Based on our findings described in Section 2,
we can expect the trace \equ(trAnx) to be
arbitrarily close to $\tr(A_\ast(x))$, if $n$ is chosen sufficiently large
and $(\alpha,E)$ sufficiently close to $(\alpha_\ast,E_\ast)$.
Then the eigenvalues of $A_n(x)$ have to cover
a nontrivial range of values near $\pm 1$, as $x$ is varied,
since the same is true for $A_\ast(x)$.

In order to determine these eigenvalues approximately,
let us use the well-known Chambers formula: if $\gcd(p,q)=1$, then
$$
\tr\bigl(\PP_q(p/q,x)\bigr)
=\EE-2\lambda^q\cos(2\pi qx)\,,
\equation(Chambers)
$$
where $\EE$ denotes the value of the left hand side for $x=(4q)^{-1}$.
Consider $\alpha=p/q$.
Choose $q=q_m$ and $p=q_{m-1}$ with $m\gg n$,
say $m-n$ constant but large.
Then $\alpha$ is the $m$-th continued fractions approximant for $\alpha_\ast$,
and $q_m\sim\alpha_\ast^{-n}$.
Presumably, we can choose $E=E_n(\lambda)$ near $E_\ast$
in such a way that $\EE=3$,
and such that $G=(\alpha,A)$ has a zero rotation number
for at least one starting point $x$.

Notice that $G$ has a nonzero rotation number $\rot(G)$
for a given $x$ if and only the trace \equ(Chambers)
takes values between $\pm 2$.
But, unless $\lambda$ is sufficiently close to $1$,
this trace is larger than $2$ for all $x$,
in which case $G$ is purely hyperbolic.
In order to avoid this, consider
taking a limit $\lambda=\lambda_n\to 1$, in such a way that
the right hand side of \equ(Chambers) approaches $2$ for $x=0$.
(Recall that $\EE=3$.)
Then
$$
1-\lambda_n\simeq-\log(\lambda_n)\simeq{\log(4/3)\over q_m}
\simeq C\alpha_\ast^n\,.
\equation(approxlambdan)
$$
This accumulation rate suggests that
$D\buR^3(P_\ast)$ has an unstable direction
with eigenvalue $\alpha_\ast^{-3}$,
related to the variation of the parameter $\lambda$ in the \AM model.

\smallskip
Another formal argument involves the fluctuations $f_n$ and $g_n$
of the rotation number $\rot(F_n)$ and $\rot(G_n)$,
respectively, around their mean values.
Here, consider a pair $P$ close to the fixed point $P_\ast$ of $\buR^3$.
Then the rotation numbers are close to zero,
and we may assume that $\rot\bigl(F_nG_n^{-1}\bigr)=\rot(F_n)-\rot(G_n)$.
Assuming furthermore that $f_ng_n$ has mean zero,
we find that the variances of $f_n$ and $g_n$ satisfy
$$
\twovec{V(f_{n+1})}{V(g_{n+1})}
=\twomat{0}{1}{1}{1}\twovec{V(f_n)}{V(g_n)}\,.
\equation(VarianceRG)
$$
Given that the matrix in this equation has an eigenvalue $\alpha_\ast^{-1}$,
this is another indication that $D\buR^3(P_\ast)$
has an eigenvalue $\alpha_\ast^{-3}$,
associated with the strength of the $x$ dependence.

\section Computer estimates

What remains to be done is to verify \clm(Contr) and \clm(AiCi).
This is carried out with the aid of a computer.
This part of the proof is written in the
programming language Ada [\rAda] and can be found in [\rFiles].
The following is meant to be a rough guide for the reader
who wishes to check the correctness of our programs.

Included in [\rFiles] are two files
{\tt approx-Fix} and {\tt ContrMat.134},
which contain the approximate fixed point $\bar P$
and the (finite rank) operator $M$, respectively,
that enter the definition \equ(ContrRGThree)
of the transformation $\buM$.

The main parts of the proof are described in the Ada package
{\tt Taylors1.Skews.Pairs}, using procedures
defined in several lower-level packages.
The main program {\tt Check\_Fixpt} first
instantiates the required packages with the appropriate parameters,
then reads $\bar P$ and $M$ from the above-mentioned files,
and finally handles control to the procedure {\tt ContrFix}
in {\tt Taylors1.Skews.Pairs}.
To give a rough idea of what happens next:
{\tt ContrFix} first computes an upper bound $\eps$ on the norm of $\buM(0)$,
and an upper bound $K$ on the norm of $D\buM(p)$ that holds
for all $p$ of norm $4\eps$ or less.
After checking that $K<3/4$, a number
$\delta<4\eps$ is chosen in such a way that $\eps+K\delta<\delta$.

These steps yield accurate and rigorous bounds on all quantities involved.
So the last statement in \clm(Contr),
as well as \clm(AiCi), are trivial to verify in this process.
In this context, a ``bound'' on a map $f:\XX\to\YY$
is a function $F$ that assigns to a set $X\subset\XX$
of a given type ({\tt Xtype}) a set $Y\subset\YY$
of a given type ({\tt Ytype}), in such a way that
$y=f(x)$ belongs to $Y$ for all $x\in X$.
In Ada, such a bound $F$ can be implemented by defining
an appropriate {\tt procedure F(X\col in Xtype\scol Y\col out Ytype)}.

Enclosures for real numbers are defined by data of type {\tt Ball}.
For common finite-dimensional spaces we use types
{\tt Vector}, {\tt Matrix}, and {\tt Polynom1}.
Our type {\tt Taylor1} provides enclosures
for functions in the spaces $\AA_\rho$.
Basic bounds for this type are defined in the package {\tt Taylors1}.
For a detailed description we refer to [\rAKfhn],
where the same type has been used.
Enclosures for matrix function in $\AA_\rho^4$
are implemented by the type {\tt Skew} defined in
the package {\tt Taylors1.Skews}.
And for pairs in $\BB_\rho$ we use a type {\tt Skew2}
defined in {\tt Taylors1.Skews.Pairs}.

Among the procedures defined in {\tt Taylors1.Skews}
is a bound {\tt Prod\_GFG} on the product
$(F,G)\mapsto GFG$ for reversible matrix functions.
Notice that the result is again reversible.
Combined with a bound {\tt Inv} on $F\mapsto F^{-1}$,
{\tt Prod\_GFG} is used to compute the composed map
$GF^{-1}G$ that appears in the first component of $\buR(P)$.
The second component involves $FG^{-1}FG^{-1}F$,
which can be computed by applying {\tt Prod\_GFG} twice.
A bound on the scaling
$(F,G)\mapsto\bigl(\Lambda_3^{-1}F\Lambda_3\,\bcomma\,\Lambda_3^{-1}G\Lambda_3\bigr)$
is defined by the procedure {\tt Equalize}.
The normalization map $\NN$ and its derivative \equ(DerNN)
are bounded via {\tt Normalize} and {\tt DNormalize}, respectively.
A bit more complex are the derivative bounds
{\tt DProd\_GFG} and {\tt DEqualize}.
But it should not be difficult to understand the code
and verify its correctness.

Bounds on the transformations $\buR_3$, $\buF$, $\buM$,
and their derivatives are obtained simply
by composing the bounds described above.

We will not explain here the more basic ideas and techniques
underlying computer-assisted proofs in analysis.
This has been done to various degrees in several other papers, including [\rAK,\rAKfhn].
As far as our proof of \clm(Contr) and \clm(AiCi) is concerned,
the ultimate reference is the source code of our programs [\rFiles].
For the center of the type {\tt Ball} we use
high precision [\rMPFR] floating-point numbers (type {\tt MPFloat}),
and for the radii we use standard [\rIEEE]
extended floating-point numbers (type {\tt LLFloat}).
Both types support controlled rounding. Our programs were run successfully on
a standard desktop machine, using a public version of the gcc/gnat compiler [\rGnat].
Instructions on how to compile and run these programs can be found in the file
{\tt README} that is included with the source code in [\rFiles].

\bigskip\noindent
{\bf Acknowledgments}.
The author would like to thank Gianni Arioli and  Sa\v sa Koci\'c
for helpful discussions, and Sa\v sa Koci\'c
for drawing my attention to the Hofstadter butterfly.

\bigskip
\references

{\ninepoint

\item{[\rHarp]} P.G.~Harper,
{\sl Single band motion of conduction electrons in a uniform magnetic field},
Proc. Phys. Soc. Lond. A {\bf 68}, 874--892 (1955).

\item{[\rHof]} D.R.~Hofstadter,
{\sl Energy levels and wave functions of Bloch electrons
in rational and irrational magnetic fields},
Phys. Rev. B 14, 2239--2249 (1976).

\item{[\rKada]} L.P.~Kadanoff,
{\it Scaling for a critical Kolmogorov--Arnold--Moser trajectory.}
Phys. Rev. Lett. {\bf 47}, 1641--1643 (1981).

\item{[\rMcK]} R.S.~MacKay,
{\it Renormalisation in Area Preserving Maps.}
Thesis, Princeton (1982).
World Scientific, London (1993).

\item{[\rBeSi]} J.~Bellissard, B.~Simon,
{\sl Cantor spectrum for the almost Mathieu equation},
J. Funct. Anal. {\bf 48}, 408--419 (1982).

\item{[\rJoMo]} R.~Johnson, J.~Moser,
{\sl The rotation number for almost periodic potentials},
Commun. Math. Phys. {\bf 84}, 403--438 (1982).

\item{[\rAvSi]} J.~Avron, B.~Simon,
{\sl Almost periodic Schr\"odinger operators. II.
The integrated density of states},
Duke Math. J. {\bf 50}, 369--391 (1983).

\item{[\rOsKi]} S.~Ostlund, S.~Kim,
{\sl Renormalization of quasiperiodic mappings},
Physica Scripta T {\bf 9}, 193--198 (1985).

\item{[\rDecJ]} C.~DeConcini, R.A.~Johnson,
{\sl The algebraic-geometric AKNS potentials},
Ergodic Theory Dynam. Syst. {\bf 7}, 1--24 (1987).

\item{[\rThou]} D.J.~Thouless,
{\sl Scaling for the discrete Mathieu equation},
Commun. Math. Phys. {\ bf 127}, 187--193 (1990).

\item{[\rMR]} M.~Rychlik, {\it
Renormalization of cocycles and linear ODE
with almost-periodic coefficients},
Invent. Math. {\bf 110}, 173--206 (1992).

\item{[\rLaWi]} Y.~Last, M.~Wilkinson,
{\sl A sum rule for the dispersion relations of the rational Harper's equation},
J. Phys. A {\bf 25}, 6123--6133 (1992).

\item{[\rLasti]} Y.~Last,
{\sl Zero measure spectrum for the almost Mathieu operator},
Comm. Math. Phys. {\bf 164}, 421--432 (1994).

\item{[\rBES]} J.~Bellissard, A.~van Elst, H.~Schulz-Baldes,
{\sl The Non-Commutative Geometry of the Quantum Hall Effect},
J. Math. Phys. {\bf 35}, 5373--5451 (1994).

\item{[\rLastii]} Y.~Last,
{\sl Almost everything about the almost Mathieu operator. I},
In: XIth International Congress of Mathematical Physics (Paris, 1994),
pp. 366--372, Cambridge MA: Internat. Press, 1995.

\item{[\rKeSa]} J.A.~Ketoja, I.I.~Satija,
{\sl Self-similarity and localization},
Phys. Rev. Lett. {\bf 75}, 2762--2765 (1995).

\item{[\rStirn]} A.~Stirnemann,
{\it Towards an Existence Proof of MacKay's Fixed Point.}
Comm.~Math. Phys. {\bf 188}, 723--735 (1997).

\item{[\rGJLS]} A.Y.~Gordon, S.~Jitomirskaya, Y.~Last, B.~Simon,
{\sl Duality and singular continuous spectrum in the almost Mathieu equation},
Acta Math. {\bf 178}, 169--183 (1997).

\item{[\rRuPi]} A.~R\"udinger, F.~Pi\'echon,
{\sl Hofstadter rules and generalized dimensions of the spectrum of Harper's equation},
J. Phys. A {\bf 30}, 117--128 (1997).

\item{[\rJito]} S.~Jitomirskaya,
{\sl Metal-insulator transition for the almost Mathieu operator},
Ann. of Math. {\bf 150}, 1159--1175 (1999).

\item{[\rMeOsWi]} B.D.~Mestel, A.H.~Osbaldestin, B.~Winn,
{\sl Golden mean renormalisation for the Harper equation:
the strong coupling fixed point},
J. Math. Phys. {\bf 41}, 8304--8330 (2000).

\item{[\rOsAv]} D.~Osadchy, J.E.~Avron,
{\sl Hofstadter butterfly as quantum phase diagram},
J. Math. Phys. {\bf 42}, 5665--5671 (2001).

\item{[\rRK]} R.~Krikorian, {\it
Global density of reducible quasi-periodic cocycles
on ${\bf T}^1\times\rmSU(2)$},
Ann. of Math. (2) {\bf 154}, 269--326 (2001).

\item{[\rMeOs]} B.D.~Mestel, A.H.~Osbaldestin,
{\sl A garden of orchids: a generalized Harper equation
at quadratic irrational frequencies}.
J. Phys. A {\bf 37}, 9071--9086 (2004).

\item{[\rAvKr]} A.~Avila, R.~Krikorian,
{\sl Reducibility or nonuniform hyperbolicity
for quasiperiodic Schr\"odin\-ger cocycles},
Ann. Math. {\bf 164}, 911--940 (2006).

\item{[\rDama]} D.~Damanik,
{\sl The spectrum of the almost Mathieu operator},
Lecture series in the CRC 701 (2008).

\item{[\rGoSch]} M. Goldstein and W. Schlag,
{\sl Fine properties of the integrated density of states
and a quantitative separation property of the Dirichlet eigenvalues},
Geom. Funct. Anal. {\bf 18}, 755--869 (2008).

\item{[\rAvDa]} A.~Avila, D.~Damanik,
{\sl Absolute continuity of the integrated density of states
for the almost Mathieu operator with non-critical coupling},
Invent. Math. {\bf 172}, 439--453 (2008).

\item{[\rAK]} G.~Arioli and H.~Koch,
The critical renormalization fixed point
for commuting pairs of area-preserving maps,
{\it Comm.~Math. Phys.}, {\bf 295} (2010), 415--429.

\item{[\rAKfhn]} G.~Arioli, H.~Koch,
{\sl Existence and stability of traveling
pulse solutions for the FitzHugh-Nagumo equation},
Nonlinear Analysis A. {\bf 113}, 51--70 (2015).

\item{[\rSatij]} I.I.~Satija,
{\sl A tale of two fractals:
the Hofstadter butterfly and the integral Apollonian gaskets},
Eur. Phys. J. Spec. Top. {\bf 225}, 2533--2547 (2016).

\item{[\rKoch]} H.~Koch,
{\sl On hyperbolicity in the renormalization of
near-critical area-preserving maps},
Discrete Contin. Dyn. Syst. {\bf 36}, 7029--7056 (2016).

\item{[\rOEIS]} Sequence
\pdfclink{0 0 1}{A049651}{https://oeis.org/A049651}
at The On-Line Encyclopedia of Integer Sequences.

\item{[\rKoKo]} H.~Koch, S.~Koci\'c,
work in progress.


\item{[\rFiles]} H.~Koch.
The source code for our programs, and data files, are available at\hfill\break
\pdfclink{0 0 1}{{\tt web.ma.utexas.edu/users/koch/papers/skewrg/}}
{http://web.ma.utexas.edu/users/koch/papers/skewrg/}

\item{[\rAda]} Ada Reference Manual, ISO/IEC 8652:2012(E),
available e.g. at\hfil\break
\pdfclink{0 0 1}{{\tt www.ada-auth.org/arm.html}}
{http://www.ada-auth.org/arm.html}

\item{[\rGnat]}
A free-software compiler for the Ada programming language,
which is part of the GNU Compiler Collection; see
\pdfclink{0 0 1}{{\tt gnu.org/software/gnat/}}{http://gnu.org/software/gnat/}

\item{[\rIEEE]} The Institute of Electrical and Electronics Engineers, Inc.,
{\sl IEEE Standard for Binary Float\-ing--Point Arithmetic},
ANSI/IEEE Std 754--2008.

\item{[\rMPFR]} The MPFR library for multiple-precision floating-point computations
with correct rounding; see
\pdfclink{0 0 1}{{\tt www.mpfr.org/}}{http://www.mpfr.org/}
}

\bye